\documentclass[acmsmall,screen]{acmart}

\AtBeginDocument{%
  \providecommand\BibTeX{{%
    \normalfont B\kern-0.5em{\scshape i\kern-0.25em b}\kern-0.8em\TeX}}}

\usepackage{fixltx2e}
\usepackage{amsmath, amsfonts}
\usepackage{graphicx}
\usepackage{algorithm}
\usepackage{algorithmicx}
\usepackage{amsmath,amsfonts}
\usepackage{array}
\usepackage{balance}
 \usepackage{booktabs}
\usepackage[skip=1pt,labelfont=bf]{caption}
 \usepackage{calligra}
\usepackage{color}
\usepackage{colortbl}
\usepackage{courier}
\usepackage{csvsimple}
\usepackage{enumitem}
\usepackage{fancybox}
\usepackage{fontenc}
\usepackage{graphicx}

\usepackage{listings}
\lstdefinelanguage{json}{
  morestring=[b]",%
  morecomment=[l]{//},
  morekeywords={true,false,null},
  stringstyle=\color{red},
  keywordstyle=\color{blue},
  commentstyle=\color{gray},
  basicstyle=\ttfamily,
}

\usepackage{longtable}
\usepackage{lscape}
\usepackage{makecell}
\usepackage{marvosym}
\usepackage{moreverb}
\usepackage{multicol}
\usepackage{multirow}
\usepackage{pifont}% http://ctan.org/pkg/pifont%\usepackage{subfigure}
\usepackage{rotating}
\usepackage{setspace}
\usepackage{subfigure}
\usepackage[most]{tcolorbox}
\usepackage{threeparttable}
\usepackage{tikz}
\usepackage[normalem]{ulem}
\usepackage{url}
\usepackage{soul}
\usepackage{wasysym}
\usepackage{xspace}
\usepackage{xcolor}

\usepackage{subcaption}

\usepackage[nameinlink]{cleveref}

\usepackage{lineno}\usepackage{pifont}
\definecolor{myred}{RGB}{200, 50, 50}
\definecolor{darkpastelred}{rgb}{0.76, 0.23, 0.13}
\definecolor{ao(english)}{rgb}{0.0, 0.5, 0.0}

\definecolor{darkpastelred}{rgb}{0.76, 0.23, 0.13}
\definecolor{ao(english)}{rgb}{0.0, 0.5, 0.0}

\definecolor{yellow}{RGB}{255,255,153}
\definecolor{grey}{RGB}{224,224,224}

\newboolean{showcomments}
\setboolean{showcomments}{true}
\ifthenelse{\boolean{showcomments}}
 { \newcommand{\mynote}[2]{
      \fbox{\bfseries\sffamily\scriptsize#1}
        {\small$\blacktriangleright$\textsf{\emph{#2}}$\blacktriangleleft$}}}
        { \newcommand{\mynote}[2]{}}
        
\definecolor{DarkOrange}{rgb}{0.8,0.3,0.0}
\definecolor{DarkCyan}{rgb}{0.0, 0.55, 0.55}
\definecolor{DarkCyel}{rgb}{1.0, 0.49, 0.0}
\definecolor{yellow-green}{rgb}{0.6, 0.8, 0.2}

\newcolumntype{?}{!{\vrule width 1pt}}

\newcommand{\highlight}[1]{\begin{tcolorbox}[leftrule=0mm,rightrule=0mm,toprule=0mm,bottomrule=0mm,left=2pt,right=2pt,top=2pt,bottom=2pt]
  #1
  \end{tcolorbox}
}

\makeatletter
\DeclareRobustCommand\onedot{\futurelet\@let@token\@onedot}
\def\@onedot{\ifx\@let@token.\else.\null\fi\xspace}

\makeatother

\begin{document}

%\title{Bridging Human and Machine: Evaluating Code Readability Using Large Language Models \tb{I am suggesting "From Metrics to Models: Large Language Models as Human-Aligned Code Readability Evaluators"}}
%ICSE rejected
%\title{From Metrics to Models: Large Language Models as Human-Aligned Code Readability Evaluators}
%For Arxiv 
\title{Human-Aligned Code Readability Assessment with Large Language Models}

\author{Wendkûuni C. Ouédraogo}
\email{wendkuuni.ouedraogo@uni.lu}
\affiliation{
  \institution{University of Luxembourg}
 	\country{Luxembourg}
}

\author{Yinghua Li}\authornote{Corresponding author.}
\email{yinghua.li@uni.lu}
\affiliation{
  \institution{University of Luxembourg}
 	\country{Luxembourg}
}

\author{Xueqi Dang}
\email{xueqi.dang@uni.lu}
\affiliation{
  \institution{University of Luxembourg}
 	\country{Luxembourg}
}

\author{Pawel BORSUKIEWICZ}
\email{pawel.borsukiewicz@uni.lu}
\affiliation{
  \institution{University of Luxembourg}
  \country{Luxembourg}
}

\author{Xin Zhou}
\email{xinzhou.2020@phdcs.smu.edu.sg}
\affiliation{
  \institution{Singapore Management University}
  \country{Singapore}
}

\author{Anil Koyuncu}
\email{anil.koyuncu@cs.bilkent.edu.tr}
\affiliation{
  \institution{Bilkent University}
 	\country{Turkey}
}

\author{Jacques Klein}
\email{jacques.klein@uni.lu}
\affiliation{
  \institution{University of Luxembourg}
  \country{Luxembourg}
}

\author{David Lo}
\email{davidlo@smu.edu.sg}
\affiliation{
  \institution{Singapore Management University}
  \country{Singapore}
}

\author{Tegawend\'e F. Bissyand\'e}
\email{tegawende.bissyande@uni.lu}
\affiliation{
  \institution{University of Luxembourg}
 	\country{Luxembourg}
}

\renewcommand{\shortauthors}{Ouédraogo and Li et al.}

\begin{abstract}

Code readability plays a pivotal role in software comprehension, maintenance, and team collaboration, yet remains difficult to assess at scale. Traditional static metrics (e.g., line length, nesting depth) often fail to capture the subjective, context-sensitive nature of human judgments. Large Language Models (LLMs) offer a scalable and interpretable alternative, but their behavior as code readability evaluators remains underexplored and insufficiently characterized. In this paper, we introduce \textbf{CoReEval}, the first large-scale benchmark for evaluating LLM-based code readability assessment. It comprises over \textbf{1.4 million} model–snippet–prompt evaluations across 10 state-of-the-art LLMs. The benchmark spans 3 programming languages (Java, Python, CUDA), 2 code types (functional code and unit tests), 4 prompting strategies (Zero-Shot (ZSL), Few-Shot (FSL), Chain-of-Thought (CoT), Tree-of-Thought (ToT)), 9 decoding settings, and developer-guided prompts tailored to junior and senior personas. We compare LLM outputs against human annotations and a validated static model, analyzing both numerical alignment (via MAE, Pearson’s $r$, and Spearman’s $\rho$) and justification quality (via sentiment, aspect coverage, and semantic clustering). Our findings show that developer-guided prompting, grounded in human-defined readability dimensions improves alignment in structured contexts (e.g., test code), enhances explanation quality, and enables lightweight personalization through persona framing. However, we also observe increased score variability, highlighting trade-offs between alignment, stability, and interpretability. CoReEval provides a robust, reproducible foundation for future work in LLM-based quality assessment. It supports prompt engineering, model alignment studies, and human-in-the-loop evaluation protocols. Beyond research, CoReEval opens promising applications in education, onboarding, and CI/CD pipelines—where LLMs can serve as explainable, adaptable reviewers aligned with developer expectations.

\end{abstract}




\maketitle

\section{Introduction}

LLMs are rapidly transforming software development, powering tools such as GitHub Copilot\footnote{\url{https://github.com/features/copilot}} and JetBrains AI Assistant\footnote{\url{https://plugins.jetbrains.com/plugin/22282-jetbrains-ai-assistant}}, which support code generation, completion, documentation, and testing. As AI-generated code becomes increasingly prevalent in production workflows, the ability to assess its quality, especially code readability has become a critical need for software teams. Readability directly affects comprehension, maintainability, and reviewability, and plays a key role in onboarding, refactoring, debugging, and team collaboration. However, readability remains inherently subjective and difficult to evaluate at scale. Traditional metrics (e.g., nesting depth, line length, or identifier entropy) often fail to capture the nuanced, developer-centered judgments that drive real-world readability assessments~\cite{buse2008metric,dorn2012general,scalabrino2018comprehensive}.

Human evaluation remains the gold standard for assessing code readability, as it aligns with developers’ cognitive processes and team conventions. However, manual assessments are time-consuming, difficult to scale, and prone to fatigue and inconsistency~\cite{kumar2024llms,zhang2024review,sergeyuk2024assessing}. This raises a key question: \textit{Can LLMs serve as reliable and interpretable readability evaluators?} LLMs combine code reasoning with natural language understanding, and thus offer a promising alternative for large-scale evaluation. While prior work has applied LLMs to tasks such as summarization and generation, their ability to replicate human readability assessments, particularly for unit tests, which follow distinct structural and stylistic patterns remains largely unexplored~\cite{winkler2024investigating,chen2024chatunitest}.

Traditional readability models rely on static code metrics such as line length, nesting depth, or identifier entropy as proxies for perceived readability~\cite{buse2008metric,dorn2012general,scalabrino2018comprehensive,mi2022towards}. However, multiple studies have shown that such features often fail to capture the nuanced and subjective criteria developers associate with readable code such as naming clarity, structural organization, and explanatory intent~\cite{sergeyuk2024reassessing,vitale2025personalized,sergeyuk2024assessing}. More fundamentally, readability is not absolute: it varies with developer expectations, team conventions, and task-specific context. Human-centric studies~\cite{sergeyuk2024assessing} have identified key developer-valued dimensions (e.g., structure, naming, clarity), but their reliance on manual annotation and context-specific rubrics limits scalability. In this study, we investigate whether LLMs, guided by developer-informed prompts, can deliver more accurate and personalized readability assessments across languages, code types, and developer profiles.

In this study, we investigate the capacity of LLMs to evaluate code readability in alignment with human expectations. To this end, we introduce \textbf{CoReEval}, a large-scale and extensible benchmark designed to assess LLMs as human-aligned readability evaluators. CoReEval spans 656 real-world code snippets both functional and test code across three programming languages (Java, Python, and CUDA), annotated with developer-assigned readability scores from widely used datasets~\cite{buse2009learning,dorn2012general,scalabrino2018comprehensive}. While Java and Python represent mainstream languages, CUDA was intentionally included to probe LLM performance on domain-specific code with non-standard syntax and structure. We evaluate 10 leading LLMs both open-source and proprietary, producing over 1.4 million model–snippet–prompt combinations. These include variations in prompting strategies (e.g.,Zero-Shot (ZSL), Few-Shot (FSL),
Chain-of-Thought (CoT), and Tree-of-Thought (ToT)), decoding parameters, and simulated developer personas. Beyond scalar score comparisons, we analyze justification quality through sentiment polarity, aspect coverage, and semantic clustering of reasoning patterns. Our analysis is guided by four research questions that examine: (i) how well LLMs align with human readability judgments, (ii) how configuration factors such as prompting and decoding influence model behavior, (iii) whether developer-guided prompting improves alignment and interpretability, and (iv) whether LLMs can support personalized readability evaluation. Rather than aiming to replace human evaluators, our goal is to better understand how LLMs behave when acting as subjective code critics—and how their outputs can be made more useful, interpretable, and aligned.

\noindent\textbf{This paper makes the following contributions:}
\begin{itemize}[leftmargin=*]

    \item \textbf{A comprehensive, multidimensional evaluation of LLMs for code readability (RQ1–RQ3):}  
    We assess ten state-of-the-art LLMs both proprietary and open-source across 656 human-annotated code snippets in Java, Python, and CUDA. Our 1.4M+ evaluations span functional and test code, multiple personas, four prompting strategies, and nine decoding settings. Beyond traditional metrics (MAE, Pearson, Spearman, Kendall), we analyze justification quality via sentiment analysis, aspect coverage, and semantic clustering, offering fine-grained insight into LLM reasoning patterns and human alignment.

    \item \textbf{An in-depth analysis of configuration effects on alignment, variability, and personalization (RQ2):}  
    We study how prompting strategies (ZSL, FSL, CoT, ToT), decoding parameters (\texttt{temperature}, \texttt{top\_p}), and persona framing (junior vs.\ senior) shape LLM scoring behavior. Our results show that configuration choices can significantly impact both alignment with human scores and justification interpretability, highlighting prompt engineering as a critical tool for practical use.

    \item \textbf{A demonstration of developer-guided prompting to enhance alignment, interpretability, and adaptability (RQ3–RQ4):}  
    Conditioning LLMs on developer-informed readability dimensions improves score alignment, expands aspect coverage, and produces more interpretable justifications, especially for test code. This strategy enables lightweight personalization without retraining, showing how LLMs can adapt to diverse team conventions and readability expectations.

    \item \textbf{CoReEval: a reproducible and extensible benchmark for future research:}  
    We introduce \textbf{CoReEval}, a large-scale benchmark for evaluating LLMs as code readability assessors. It comprises over \textbf{1.4M model–snippet evaluations}, each including a readability score and rationale, spanning 10 models, 3 programming languages, 2 code types, 4 prompting strategies, and 9 decoding settings. For interpretability analysis, we extract and analyze more than \textbf{10K representative justifications}, covering diverse reasoning patterns and aspect combinations. All datasets, prompt templates, human annotations, and analysis scripts are released in an open replication package\footnote{\url{https://anonymous.4open.science/r/CoReEval-072D/}}—enabling reproducible research in LLM alignment, automated code review, education, and CI/CD quality tooling.

\end{itemize}

The paper is organized as follows: \Cref{background} outlines key concept. \Cref{approach} details our experimental setup. \Cref{results} presents findings for the four research questions. \Cref{discussion} discusses implications and limitations. \Cref{relatedwork} positions our study within existing literature, and \Cref{conclusion} concludes with key insights and future directions.

\section{Background}
\label{background}

\subsection{Large Language Models}
Driven by natural language prompts~\cite{siddiq2024using,vogelsang2024using}, large language models (LLMs) have become increasingly effective across a wide range of software engineering tasks, including code generation~\cite{chen2021evaluating,li2023starcoder}, test synthesis~\cite{chen2023chatunitest}, and documentation~\cite{fan2023large}. Tools such as GitHub Copilot and OpenAI Codex\footnote{\url{https://openai.com/index/openai-codex/}} exemplify the growing integration of LLMs into real-world developer workflows. Beyond generation, LLMs are now being explored as evaluators of software artifacts, such as in code summarization~\cite{sun2024source}, bug understanding~\cite{kumar2024llms}, and code quality scoring~\cite{wang2025can}. Wang et al.~\cite{wang2025can} notably study LLMs as automated judges, showing that while prompting strategies like Chain-of-Thought (CoT) and majority voting improve performance, model judgments still exhibit variability and depend heavily on task context.

Recent studies have also highlighted the influence of two critical factors on LLM evaluation behavior. First, the \textbf{persona framing} used in prompts can shape both reasoning style and outputs. Kong et al.~\cite{kong2024better} show that role-playing prompts (e.g., "as a software architect") improve reasoning in zero-shot settings, while Dong et al.~\cite{dong2024self} demonstrate that developer roles (e.g., QA engineer vs. backend engineer) impact generated code and evaluation judgments. These findings motivate the exploration of developer personas (junior vs.\ senior) in this study. Second, model \textbf{scale} (i.e., parameter size) has emerged as a key determinant of LLM capabilities. Larger models (e.g., 64B+) tend to produce more coherent outputs and better align with human expectations~\cite{achiam2023gpt,dong2024self,wang2025can,sun2024source}, whereas smaller open-source models (4–7B) offer reduced latency and greater deployment flexibility, but with trade-offs in reasoning quality and consistency~\cite{wang2025can, naveed2023comprehensive}. Understanding these trade-offs is essential when considering LLMs as evaluators in practical settings.

Building on these developments, our study investigates the viability of LLMs as readability evaluators—i.e., as automated judges that assign human-aligned scores and justifications to code. We aim to bridge the gap between scalable automation and human-centered quality assessment, across both functional and test code in multiple programming languages.

%\vspace{-2mm}
\subsection{Source Code Quality}
Source code quality spans attributes such as maintainability, testability, understandability, and readability. While traditional metrics—like Lines of Code (LOC), Cyclomatic Complexity~\cite{mccabe1976complexity}, Cognitive Complexity~\cite{sharma2020we}, and Halstead metrics~\cite{lavazza2023empirical}—offer structural insights, they often fall short in capturing subjective, human-centric dimensions~\cite{dantas2021readability,lavazza2023empirical}. Recent efforts seek to bridge this gap. Cognitive Complexity incorporates aspects of readability~\cite{sharma2020we}, while approaches like that of Papamichail et al.~\cite{papamichail2016user} model user-perceived quality from static metrics and GitHub data. However, empirical studies~\cite{lavazza2023empirical} highlight persistent challenges in accurately predicting understandability. LLMs offer a promising alternative. By processing both syntactic and semantic signals, they enable context-aware suggestions that support code generation, refactoring, and comprehension~\cite{fan2023large,wadhwa2023frustrated,chen2021evaluating}. 
This study centers on code readability as a key pillar of source code quality. We investigate whether LLMs can perform reliable, interpretable, and context-sensitive readability assessments that align with how developers evaluate code in practice.

%\vspace{-2mm}
\subsection{Readability vs Understandability}
Readability and understandability are closely related yet distinct dimensions of code quality. Readability refers to how easily code can be visually parsed—driven by layout, indentation, and naming~\cite{buse2009learning,oliveira2022systematic}—while understandability concerns the developer’s ability to grasp the code’s intent, logic, and behavior~\cite{grano2018empirical,winkler2024investigating}. Traditional readability models~\cite{buse2009learning,scalabrino2018comprehensive,mi2022towards} emphasize visual and structural features, with recent efforts extending to unit test code~\cite{daka2015modeling}. In contrast, understandability is often assessed using metrics like cyclomatic or cognitive complexity~\cite{mccabe1976complexity,campbell2018cognitive}, which may overlook test-specific issues such as verbose assertions or fixture entanglement~\cite{winkler2024investigating,guerra2024annotations}. 
While readability supports surface-level navigation and understandability enables deeper reasoning, both are intertwined—especially in test code~\cite{scalabrino2018comprehensive,sergeyuk2024reassessing}. We focus on \textit{readability} as perceived by developers, while examining whether LLMs—guided by prompts—can generate assessments that subtly capture deeper cues relevant to understanding, without shifting the evaluation scope.

\section{Study design}
\label{approach}

\subsection{Analysis Overview}
\label{pipeline_overview}
\begin{figure*}[t]
\centering
\scalebox{0.35}{\includegraphics{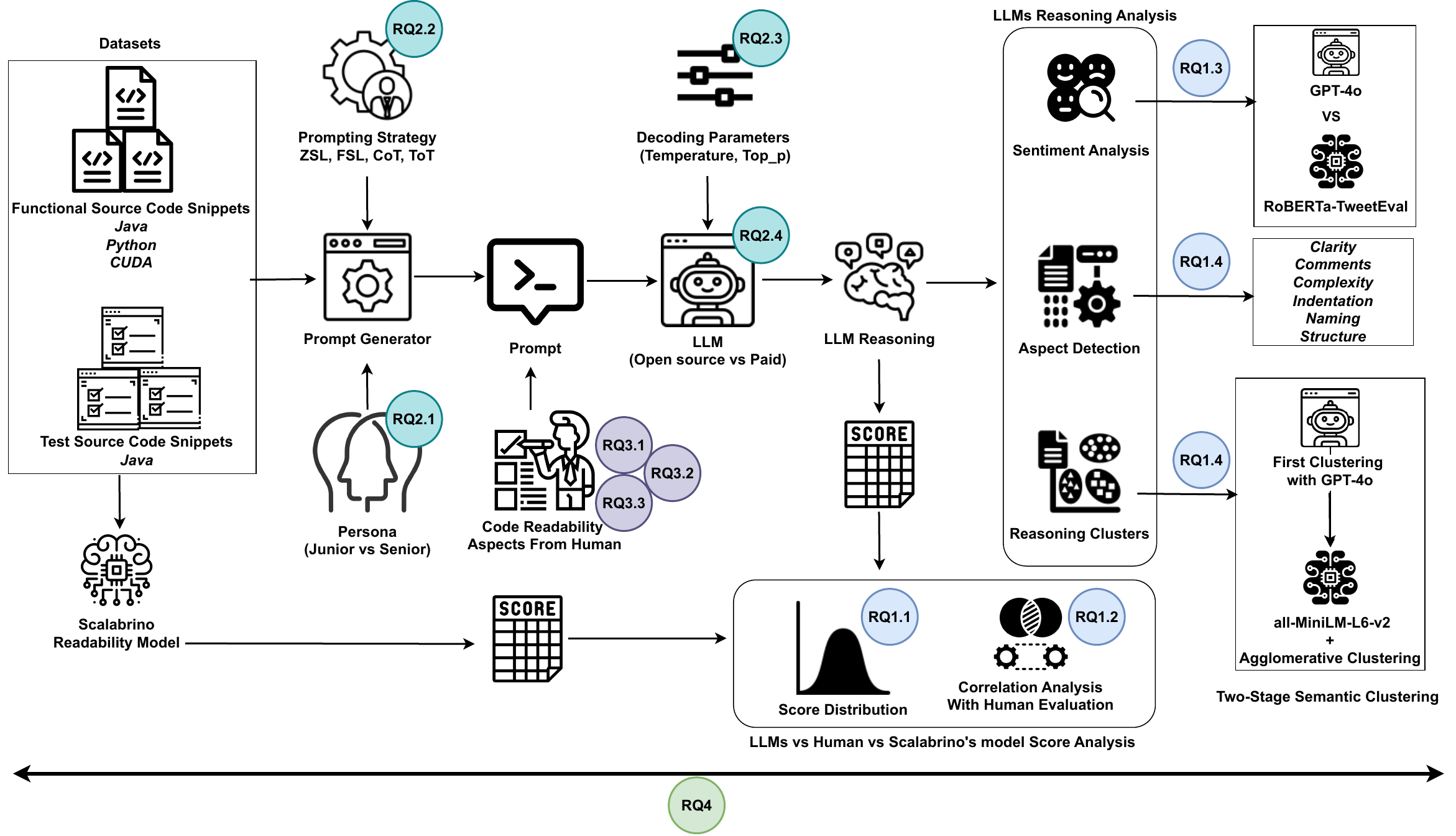}}
%\vspace{1mm}
\caption{Overview of our explorative study.}
% \vspace{2mm}
\label{fig:overview}
\end{figure*}

This study adopts a structured, multi-phase methodology to evaluate how LLMs assess source code readability. Our pipeline covers diverse languages (Java, Python, CUDA), code types (functional and test), and model configurations (prompts, personas, decoding parameters). Guided by four research questions (RQs), we analyze LLM outputs in terms of score alignment, justification quality, sensitivity to prompting and decoding parameter factors, and potential for personalized evaluation. Figure~\ref{fig:overview} provides a high-level summary of the process.

%%%%%%%%%%%%%%%%%%%%%%%%%%%%%%%%%%%%##############
\subsection{Research Questions}
\label{sub:RQ}
Our experimental evaluation is structured around four research questions that progressively assess the capabilities and limitations of LLMs as human-aligned code readability evaluators.

\vspace{0.2cm}\noindent\textbf{RQ1: How well do LLMs align with human judgments of code readability?}  
This first question establishes a baseline comparison between LLM-generated readability assessments and human ratings across both functional and unit test code. We evaluate alignment using quantitative metrics such as Mean Absolute Error (MAE), Pearson's $r$, and Spearman's $\rho$. Additionally, we analyze the content of model justifications through sentiment balance, aspect coverage, and semantic clustering. For Java code, we include direct comparisons with Scalabrino et al.'s static readability model~\cite{scalabrino2018comprehensive}.

\vspace{0.2cm}\noindent\textbf{RQ2: What factors influence LLM performance in readability evaluation tasks?}  
We investigate how various configuration parameters impact both the scoring behavior and justification quality of LLMs. Specifically, we analyze the role of developer persona (junior vs.\ senior), prompting strategy (ZSL, FSL, CoT, ToT), decoding parameters (temperature and top\_p), and model scale (small vs.\ large). This allows us to identify which factors most affect evaluation reliability and interpretability.

\vspace{0.2cm}\noindent\textbf{RQ3: Does developer-guided prompting improve the quality of LLM readability evaluations?}  
Building on RQ2, this question investigates whether enriching prompts with structured, human-centric evaluation criteria improves the quality of LLM-based readability assessments. We evaluate gains in score accuracy, alignment with human judgments, and consistency across code types and languages. For Java code, we include direct comparisons with Scalabrino et al.'s static readability model~\cite{scalabrino2018comprehensive} to contextualize improvements. We also analyze justification depth, focusing on the coverage of developer-relevant aspects.

\vspace{0.2cm}\noindent\textbf{RQ4: Can developer guidance support personalized readability evaluation with LLMs?}  
Finally, we synthesize results from the previous questions to assess whether LLMs can adapt their evaluations to different user profiles and code contexts. We focus on whether developer-guided prompting, when combined with persona framing, enables context-aware and subjective yet consistent readability judgments, paving the way for personalized tooling in development workflows.

%\vspace{-4mm}
\subsection{Models}
\label{subsec:models}
To benchmark the effectiveness of different LLMs in source code readability evaluation, we include a diverse set of 10 models that vary in architecture, size, and provider. These include both proprietary and open-source models, covering a wide spectrum of capabilities and inference modalities.

We categorize the evaluated models by their \textbf{model scale}, as prior studies have shown that model size plays a critical role in shaping LLM reasoning, stability, and evaluation reliability in software engineering contexts~\cite{wang2025can, dong2024self}. In line with these findings, we distinguish between \emph{small-scale models} (4–7B parameters) and \emph{large-scale models} (64B+ parameters), enabling more meaningful comparisons across capacity classes while controlling for architectural and provider-related variability.

Table~\ref{tab:llms-used} lists the models grouped by size category. To ensure a fair and reproducible evaluation, we systematically vary decoding parameters across all models. Following the protocol of Sun et al.~\cite{sun2024source}, we use a 3x3 grid of \texttt{temperature} and \texttt{top\_p} values (Table~\ref{tab:decoding-settings}) to assess sampling stability and generation diversity.

\begin{table}[ht]
\centering
\caption{Large Language Models evaluated.}
\label{tab:llms-used}
\scalebox{0.8}{
\begin{tabular}{llll}
\toprule
\multicolumn{4}{c}{\textbf{API-based Models}} \\
\textbf{Model} & \textbf{Provider} & \textbf{Size} & \textbf{Access} \\
\midrule
Claude 3 Sonnet       & Anthropic     & Unknown         & API   \\
Deepseek Chat         & Deepseek      & $\sim$64B       & API   \\
GPT-4o                & OpenAI        & Unknown         & API   \\
Mistral Large (v2407) & Mistral       & $\sim$65B (MoE) & API   \\
\midrule
\multicolumn{4}{c}{\textbf{Local Models}} \\
\textbf{Model} & \textbf{Provider} & \textbf{Size} & \textbf{Access} \\
\midrule
Deepseek LLM 7B Chat       & Deepseek (HF)   & 7B    & Local \\
Falcon 3 7B Instruct       & TII (HF)        & 7B    & Local \\
Gemma 7B IT                & Google (HF)     & 7B    & Local \\
Mistral 7B Instruct (v0.3) & Mistral (HF)    & 7B    & Local \\
Phi-3 Mini                 & Microsoft (HF)  & 4.2B  & Local \\
Qwen2.5 7B Instruct        & Alibaba (HF)    & 7B    & Local \\
\bottomrule
\end{tabular}
}
\end{table}

%%%%%%%%%%%%%
\begin{table}[ht]
\centering
\caption{Decoding parameter combinations used.}
\label{tab:decoding-settings}
\scalebox{0.8}{
\begin{tabular}{l|ccc|ccc|ccc}
\toprule
\textbf{Temperature} & \multicolumn{3}{c|}{0.1} & \multicolumn{3}{c|}{0.5} & \multicolumn{3}{c}{1.0} \\
\textbf{Top\_p}      & 0.5 & 0.75 & 1.0         & 0.5 & 0.75 & 1.0         & 0.5 & 0.75 & 1.0 \\
\bottomrule
\end{tabular}
}
\end{table}

\subsection{Human-Centric Readability Aspects}
\label{subsec:readability-aspects}
To support developer-guided prompts, we incorporate readability dimensions derived from prior empirical studies. For functional code, we adopt the six aspects with the strongest inter-rater agreement and correlation with human readability judgments, as identified by Sergeyuk et al.\cite{sergeyuk2024assessing}. For unit test code, we follow the six most consistently cited factors across both scientific and practitioner sources, validated by Winkler et al.\cite{winkler2024investigating}. These dimensions, summarized in Table~\ref{tab:readability-aspects}, form the foundation of our prompt design in RQ3 and our personalization analysis in RQ4.

\vspace{0.5em}
\begin{table}[ht]
\centering
\caption{Top 6 human-centric readability aspects used.}
\label{tab:readability-aspects}
\scalebox{0.9}{
\begin{tabular}{p{10.cm}}
\toprule
\textbf{Functional Code (Top Developer Agreement)} \\
\midrule
\textbf{Code Structure:} Logical separation of functionality vs. tangled code \\
\textbf{Nesting:} Flat, linear code vs. deeply nested blocks \\
\textbf{Understandable Goal:} Clear task or function vs. ambiguous purpose \\
\textbf{Code Length:} Concise and readable vs. unnecessarily long code \\
\textbf{Inline Actions:} One action per line vs. multiple actions in the same line \\
\textbf{Reading Flow:} Natural, top-to-bottom logic vs. disorganized flow \\
\midrule
\textbf{Unit Test Code (Scientific + Practitioner Consensus)} \\
\midrule
\textbf{Test Structure:} Use of Arrange-Act-Assert pattern \\
\textbf{Test Names:} Descriptive and standardized naming \\
\textbf{Assertions:} Readable and well-placed test assertions \\
\textbf{Test Code Size:} Compact body avoiding verbosity \\
\textbf{Test Purpose Clarity:} Each test expresses a distinct intent \\
\textbf{Fixture Simplicity:} Setup code is lightweight and understandable \\
\bottomrule
\end{tabular}
}
\end{table}

\subsection{Prompting Techniques}
\label{sub:prompting_techniques}

To systematically explore the influence of prompting design on LLM readability assessment, we construct a comprehensive prompt matrix that varies along three orthogonal dimensions: (1) \textbf{prompt type} (generic vs.\ developer-guided), (2) \textbf{persona framing} (junior vs.\ senior), and (3) \textbf{code category} (functional vs.\ test code). Each prompt produces a structured JSON response containing a \texttt{score} between 0 and 20 and a concise \texttt{reasoning} field limited to 450 characters. This design results in a total of \textbf{64 unique prompt configurations}: 48 for functional code (spanning Java, Python, and CUDA) and 16 for test code (restricted to Java). The construction of this prompt matrix is informed by recent work on role-based prompting~\cite{kong2024better, dong2024self} and developer-centered evaluation criteria~\cite{sergeyuk2024assessing, winkler2024investigating}, ensuring a balanced and reproducible exploration of LLM behavior across diverse prompting conditions.

\textbf{Generic prompts} provide light but structured contextual framing, relying on broadly accepted readability criteria such as naming clarity, code structure, indentation, comments, and overall legibility. These criteria reflect standard best practices in software engineering and are commonly cited in both educational and industrial contexts. As such, these prompts simulate a typical evaluation scenario in which models operate based on general programming conventions, without being guided by task-specific evaluation dimensions. In contrast, \textbf{developer-guided prompts} embed the top 6 explicit readability criteria derived from prior work on human-centered code evaluation. For functional code, these criteria include code structure, nesting depth, intent clarity, code length, action granularity, and reading flow~\cite{sergeyuk2024assessing}. For test code, they reflect practices such as Arrange-Act-Assert structure, test naming clarity, assertion readability, code conciseness, test purpose clarity, and fixture simplicity~\cite{winkler2024investigating}. This setup allows us to examine whether LLMs behave more consistently or more insightfully when provided with structured, developer-informed evaluation axes.

To explore potential personalization effects, each prompt is instantiated under one of two persona framings. The \textbf{junior persona} is cast as a software engineering student focusing on code quality, while the \textbf{senior persona} reflects an experienced developer with evaluation expertise. Although both use the same information, their wording and tone differ subtly to reflect differences in expertise and perspective. This variation enables us to investigate whether LLMs exhibit framing-sensitive behavior when prompted with different evaluator profiles, as suggested in recent prompting literature~\cite{kong2024better, dong2024self}.

%\paragraph{Code Categories and Prompt Variants.}
We generate prompts for two categories of code: general-purpose functional code (across Java, Python, and CUDA), and test code (restricted to Java). The prompts are adapted to each category, with test-oriented variants incorporating domain-specific terminology such as test fixtures, assertion structure, and Arrange-Act-Assert patterns, in order to better reflect developer conventions. All prompt types and persona combinations are evaluated under four prompting strategies: \textbf{Zero-Shot (ZSL)}, which poses a direct readability question without examples; \textbf{Few-Shot (FSL)}, which provides illustrative examples of code snippets with associated scores and justifications; \textbf{Chain-of-Thought (CoT)}~\cite{wei2022chain}, which encourages step-by-step reasoning across evaluation dimensions; and \textbf{Tree-of-Thought (ToT)}~\cite{long2023large}, which simulates collaborative reasoning among multiple virtual experts leading to a final consensus. This orthogonal variation of code category, strategy, persona, and prompt type enables a comprehensive investigation of how prompting design shapes LLM behavior in readability evaluation.

%\paragraph{Prompt Examples.}
Tables~\ref{tab:prompt-examples-junior} and~\ref{tab:prompt-examples-senior} present representative prompt excerpts across the four prompting strategies (ZSL, FSL, CoT, ToT) under the junior and senior persona framings, respectively. These excerpts correspond to the Java functional code setting, where LLMs evaluate general-purpose source code. The same prompt structure and variation principles apply consistently to other programming languages (Python and CUDA) and to test-related prompts, with minor terminology adaptations (e.g., test fixtures, assertions). Overall, the tables illustrate how generic prompts rely on broadly defined readability dimensions, whereas developer-guided prompts incorporate explicit, human-aligned evaluation criteria. Full prompt templates for all combinations of persona, language, and code category are available in our replication package.

\begin{table*}[ht]
\centering
\caption{Java prompt examples for the \textbf{junior persona} across prompting strategies (generic vs. guided).}
\label{tab:prompt-examples-junior}
\scalebox{0.9}{
\begin{tabular}{p{0.11\linewidth} p{0.43\linewidth} p{0.43\linewidth}}
\toprule
\textbf{Strategy} & \textbf{Generic Prompt (Excerpt)} & \textbf{Developer-Guided Prompt (Excerpt)} \\
\midrule
ZSL &
\parbox[t]{\linewidth}{\small\texttt{You are a software engineering student specializing in code quality assessment and readability evaluation. Evaluate provided Java code on a scale from 0 (unreadable) to 20 (highly readable) based on naming clarity, structure, indentation, comments, complexity, and overall clarity. Respond ONLY with JSON: \{"score": "0-20", "reasoning": "concise justification, max 450 chars"\} [...]}} &
\parbox[t]{\linewidth}{\small\texttt{You are a software engineering student evaluating Java code readability explicitly using these human-centric criteria: 1. Code Structure [...] 6. Reading Flow. Evaluate readability (0=unreadable, 20=highly readable). Respond ONLY with JSON: \{"score": "0-20", "reasoning": "explicit justification, max 450 chars"\} [...]}} \\
\midrule
FSL &
\parbox[t]{\linewidth}{\small\texttt{You are a software engineering student specializing in code quality assessment and readability evaluation. Evaluate provided Java code on a scale from 0 to 20 based on naming, structure, indentation, comments, complexity, clarity. [...] Now, evaluate the following code: SOURCE\_CODE [...]}} &
\parbox[t]{\linewidth}{\small\texttt{You are a software engineering student evaluating Java code readability using these human-centric criteria: 1. Structure [...] 6. Reading Flow. Example 1: [...] Evaluation: \{"score": "17", "reasoning": "Clear goal, good structure..." \} [...] Now evaluate: SOURCE\_CODE}} \\
\midrule
CoT &
\parbox[t]{\linewidth}{\small\texttt{You are a software engineering student specializing in code quality assessment. Evaluate Java source code by reasoning step-by-step: Step 1: Naming clarity... Step 5: Complexity. Then provide JSON with: \{"score": "0-20", "reasoning": "..."\}}} &
\parbox[t]{\linewidth}{\small\texttt{You are a software engineering student evaluating Java code step-by-step using criteria: 1. Structure [...] 6. Reading Flow. After stepwise reasoning, provide final score. Respond ONLY with JSON: \{"score": "...", "reasoning": "..."\}}} \\
\midrule
ToT &
\parbox[t]{\linewidth}{\small\texttt{You facilitate a panel of 3 software engineering students. Each evaluates readability based on naming, structure, indentation, clarity... then reach a consensus. Provide JSON: \{"score": "...", "reasoning": "..."\}}} &
\parbox[t]{\linewidth}{\small\texttt{You facilitate 3 students evaluating readability explicitly using: 1. Structure [...] 6. Reading Flow. Students agree on a final score and justification. Provide JSON: \{"score": "...", "reasoning": "..."\}}} \\
\bottomrule
\end{tabular}
}
\end{table*}

\begin{table*}[ht]
\centering
\caption{Java prompt examples for the \textbf{senior persona} across prompting strategies (generic vs. guided).}

\label{tab:prompt-examples-senior}
\scalebox{0.9}{
\begin{tabular}{p{0.11\linewidth} p{0.43\linewidth} p{0.43\linewidth}}
\toprule
\textbf{Strategy} & \textbf{Generic Prompt (Excerpt)} & \textbf{Developer-Guided Prompt (Excerpt)} \\
\midrule
ZSL &
\parbox[t]{\linewidth}{\small\texttt{You are an experienced senior software engineer specialized in code quality assessment and readability evaluation. Evaluate provided Java code on a scale from 0 (unreadable) to 20 (highly readable) based on naming clarity, structure, indentation, comments, complexity, and overall clarity. Respond ONLY with JSON: \{"score": "0-20", "reasoning": "..."\}}} &
\parbox[t]{\linewidth}{\small\texttt{You are an experienced senior software engineer evaluating Java code readability explicitly using these criteria: 1. Code Structure [...] 6. Reading Flow. Evaluate readability (0=unreadable, 20=highly readable). Respond ONLY with JSON: \{"score": "0-20", "reasoning": "..."\}}} \\
\midrule
FSL &
\parbox[t]{\linewidth}{\small\texttt{You are an experienced senior software engineer specialized in code quality assessment and readability evaluation. [...] Here are some examples of Java source code readability evaluations: [...] Now, evaluate the following Java source code: SOURCE\_CODE [...]}} &
\parbox[t]{\linewidth}{\small\texttt{You are an experienced senior software engineer evaluating Java code readability using explicit criteria. Example 1: [...] Evaluation: \{"score": "17", "reasoning": "Clear goal, good structure..." \} [...] Now evaluate: SOURCE\_CODE}} \\
\midrule
CoT &
\parbox[t]{\linewidth}{\small\texttt{You are an experienced senior software engineer. Evaluate Java source code by reasoning step-by-step: Step 1: Naming clarity [...] Step 5: Complexity. Then provide final score using: \{"score": "...", "reasoning": "..."\}}} &
\parbox[t]{\linewidth}{\small\texttt{You are an experienced senior software engineer evaluating Java readability step-by-step using: 1. Code Structure [...] 6. Reading Flow. After reasoning, respond ONLY with JSON: \{"score": "...", "reasoning": "..."\}}} \\
\midrule
ToT &
\parbox[t]{\linewidth}{\small\texttt{You are facilitating a panel of three senior software engineers (Expert A, B, C). Each independently evaluates readability (0–20) based on naming, structure, indentation, comments, complexity, and clarity, then reaches a consensus. Final evaluation in JSON: \{"score": "...", "reasoning": "..."\}}} &
\parbox[t]{\linewidth}{\small\texttt{You facilitate 3 senior engineers evaluating Java readability using: 1. Code Structure [...] 6. Reading Flow. They discuss, resolve discrepancies, and agree on consensus score. Provide final JSON: \{"score": "...", "reasoning": "..."\}}} \\
\bottomrule
\end{tabular}
}
\end{table*}

\subsection{Datasets}
\label{sub:datasets}

To evaluate LLM-based readability assessment, we leverage three well-established datasets commonly used in the literature: Buse and Weimer~\cite{buse2009learning}, Dorn et al.\cite{dorn2012general}, and Scalabrino et al.\cite{scalabrino2018comprehensive}. These datasets (Table~\ref{tab:datasets-overview}) collectively span three programming languages (Java, Python, CUDA), cover both functional and unit test code, and provide human-annotated readability scores on a 5-point Likert scale. The \textbf{Buse and Weimer} dataset includes 100 short Java snippets, manually crafted to contain three imperative statements (e.g., assignments, calls), and almost entirely composed of functional code. Each snippet was rated by 120 computer science students, resulting in a dense evaluation matrix. While highly controlled and simplified, this dataset offers a baseline for readability modeling. The \textbf{Dorn} dataset improves generalizability by collecting 360 real-world snippets from open-source Java, Python, and CUDA projects, including both functional (325) and unit test code (35). Ratings were obtained from over 5,000 developers, forming a sparse annotation matrix. This dataset extends prior efforts by incorporating diverse code structures and additional semantic features. Finally, the \textbf{Scalabrino} dataset focuses on method-level Java code, with 200 snippets (124 functional, 76 test) drawn from real projects. Each was rated by 9 computer science students, and the dataset includes rich static metrics and test-specific attributes (e.g., assertion patterns), making it especially relevant for our analysis of human-centric readability prompts.

\begin{table}[ht]
\centering
\caption{Overview of the datasets used in this study.}
\label{tab:datasets-overview}
\scalebox{0.7}{
\begin{tabular}{llc|c|ccc|cc}
\toprule
\textbf{Dataset} & \textbf{Lang.} & \textbf{Type} & \textbf{\#Snip} & 
\makecell{\textbf{LOC}\\\textbf{avg}} & 
\makecell{\textbf{LOC}\\\textbf{max}} &
\makecell{\textbf{Cyc}\\\textbf{avg}} & 
\makecell{\textbf{Cyc}\\\textbf{max}} \\
\midrule
\multirow{2}{*}{\textbf{Buse}} 
  & Java & Code & 99 & \multirow{2}{*}{5.4} & \multirow{2}{*}{12} & \multirow{2}{*}{1.02} & \multirow{2}{*}{2} \\
  &      & Test & 1  &                      &                     &                       &                     \\
\midrule
\multirow{4}{*}{\textbf{Dorn}} 
  & Java   & Code & 90  & \multirow{4}{*}{21.71} & \multirow{4}{*}{48} & \multirow{4}{*}{2.05} & \multirow{4}{*}{15} \\
  & Java   & Test & 31  &                        &                     &                       &                      \\
  & Python & Code & 118 &                        &                     &                       &                      \\
  & CUDA   & Code & 117 &                        &                     &                       &                      \\
\midrule
\multirow{2}{*}{\textbf{Scalabrino}} 
  & Java & Code & 124 & \multirow{2}{*}{22.31} & \multirow{2}{*}{42} & \multirow{2}{*}{3.67} & \multirow{2}{*}{38} \\
  &      & Test & 76  &                        &                     &                       &                      \\
\bottomrule
\end{tabular}
}
\end{table}

%%%%%%%%%
\subsection{Implementation and Configuration}
\label{sub:implementation}
We evaluate 10 LLMs using both API-based models (e.g., GPT-4o, Claude 3 Sonnet, Mistral Large v2407) and open-source instruct variants from Hugging Face. API inference was run on a 48-core AMD EPYC server with 640,GB RAM, while open-source models were deployed on an high-performance computing (HPC) cluster with 4$\times$~Tesla V100 GPUs and 28 CPU threads per job, using the \texttt{transformers} library. Each model was tested across 9 decoding parameters (\texttt{temperature} $\in$ {0.1, 0.5, 1.0}, \texttt{top\_p} $\in$ {0.5, 0.75, 1.0}), with 3 repetitions per setting to account for stochasticity. We systematically varied the prompting strategy, developer-guided prompt presence, and evaluator persona (junior vs.\ senior), across functional and unit test code in three languages (Java, Python, CUDA), totaling 656 code snippets and \textbf{over 1.4 million LLM-based evaluations}. Sentiment classification used \texttt{RoBERTa-TweetEval}\footnote{\url{https://huggingface.co/cardiffnlp/twitter-roberta-base-sentiment-latest}} on a local workstation with an RTX 5000 GPU. LLM-based reasoning clustering was performed using GPT-4o with fixed decoding settings (\texttt{temperature}=0, \texttt{top\_p}=1). To reduce redundancy, we applied semantic reclustering with \texttt{all-MiniLM-L6-v2}\footnote{\url{https://huggingface.co/sentence-transformers/all-MiniLM-L6-v2}} embeddings and agglomerative clustering. All preprocessing and postprocessing were conducted on the same AMD server used for API inference.

\subsection{Metrics and Evaluation}
\label{sub:metrics}

To evaluate the effectiveness of LLMs in assessing code readability, we adopt a multi-faceted evaluation protocol that combines score-level accuracy with justification-level interpretability. Our metrics are aligned with recent standards in LLM-based evaluation~\cite{wang2025can, sun2024source}, and are designed to capture not only how well LLM outputs align with human judgments, but also how plausibly and transparently those outputs are justified. This section details the metrics and the prompting strategies used to extract interpretable insights.

\paragraph{Score Alignment.}  
We compute the Mean Absolute Error (MAE) between model-generated readability scores and human-annotated ground truth, and report both Pearson’s $r$ and Spearman’s $\rho$ to assess linear and monotonic correlations. These metrics follow the same evaluation strategy used in recent studies comparing LLMs and human evaluators~\cite{wang2025can, sun2024source}.  
In addition, we apply Mann–Whitney U tests to assess whether observed differences in score distributions across model configurations or prompting strategies are statistically significant. This non-parametric test complements correlation-based metrics by quantifying distributional shifts in model behavior, helping to identify cases where guidance or model scale alters the variability of readability judgments.

\paragraph{Aspect Coverage.}  
To quantify interpretability, we analyze the coverage of readability aspects in model justifications. Coverage is defined as the explicit mention of human-relevant dimensions such as naming clarity, structural organization, indentation, complexity, and overall comprehension—grounded in established readability frameworks~\cite{sergeyuk2024assessing, winkler2024investigating}. We use GPT-4o to perform binary aspect tagging via structured prompting. The following prompt was used to extract aspect-level annotations:

{\scriptsize
\begin{tcolorbox}[colback=gray!5, colframe=gray!40, title=\textbf{Prompt: Aspect Coverage Extraction},left=2pt, right=2pt, top=2pt, bottom=2pt, boxrule=0.4pt, breakable]
You are a software engineering researcher.

Analyze the following test readability reasoning and identify whether the author explicitly \textbf{mentions} each of the following aspects:

\begin{itemize}
\item naming
\item comments
\item indentation
\item structure
\item complexity
\item clarity
\end{itemize}

Return a JSON object like this, where 1 means "explicitly mentioned", 0 means "not mentioned":
\begin{lstlisting}[language=json]
{
  "naming": 0 or 1,
  "comments": 0 or 1,
  "indentation": 0 or 1,
  "structure": 0 or 1,
  "complexity": 0 or 1,
  "clarity": 0 or 1
}
Reasoning:
{reasoning}
\end{lstlisting}
\end{tcolorbox}
}

\paragraph{Sentiment Analysis.}  
To assess the evaluative tone of LLM justifications, we perform sentiment classification along a three-point polarity axis: \textit{positive}, \textit{neutral}, or \textit{negative}. We use two methods: (i) a fine-tuned transformer model (RoBERTa-TweetEval), and (ii) a structured GPT-4o prompt.

The GPT-4o prompt used for this task is shown below:

\vspace{0.3em}

{\scriptsize
\begin{tcolorbox}[colback=gray!5, colframe=gray!40, title=\textbf{Prompt: Sentiment Classification},left=2pt, right=2pt, top=2pt, bottom=2pt, boxrule=0.4pt, breakable]
Is the following feedback overall positive, neutral, or negative?

Return a JSON object with:
\begin{lstlisting}[language=json]
{
  "label": "positive|neutral|negative",
  "justification": "..."
}

Feedback:
{reasoning}
\end{lstlisting}
Return only the JSON object inside triple backticks.
\end{tcolorbox}
}

\paragraph{Semantic Clustering of Justifications.}  
To analyze the diversity of reasoning strategies used by LLMs, we conduct a two-stage clustering process. First, we extract a short summary label for each justification using GPT-4o. Then, we perform semantic reclustering over MiniLM embeddings using agglomerative clustering to discover canonical reasoning patterns.

The initial labeling is obtained using the following GPT-4o prompt:

\vspace{0.3em}

{\scriptsize
\begin{tcolorbox}[colback=gray!5, colframe=gray!40, title=\textbf{Prompt: Explanation Labeling for Clustering},left=2pt, right=2pt, top=2pt, bottom=2pt, boxrule=0.4pt, breakable]
You are a software engineering researcher.

Given the following readability evaluation reasoning, summarize it into a short label or theme (max 5 words) capturing its main idea.

Return a JSON object like:
\begin{lstlisting}[language=json]
{
  "label": "..."
}
Reasoning:
{reasoning}
\end{lstlisting}
Return only the JSON object inside triple backticks.
\end{tcolorbox}
}

\paragraph{Evaluation Philosophy.}  
Altogether, our evaluation framework combines \textit{numerical alignment} with \textit{interpretability-driven analysis}, enabling a richer and more human-centered understanding of how LLMs operate as readability evaluators. This dual focus reflects emerging best practices in evaluating LLMs across both summarization and judgment tasks in software engineering~\cite{wang2025can, sun2024source}.

\section{Results and analysis}
\label{results} 

\subsection{RQ1: Alignment between LLM and human readability assessments}
\label{subsec:RQ1}

\noindent \textbf{[Experimental design]:} 
This research question evaluates how closely LLM-based readability assessments align with human judgments—both in terms of numeric scores and justification quality. We assess 656 code snippets spanning three languages (Java, Python, CUDA) and two code types (functional and test), under generic prompting. LLM scores are normalized to a 1–5 Likert scale and compared to human annotations using Pearson, Spearman, and Kendall correlation coefficients. For Java, we also benchmark against Scalabrino’s model as a static baseline. To evaluate explanation quality, we perform dual sentiment analysis using RoBERTa-TweetEval and GPT-4o. We extract semantic aspects from GPT-4o rationales and conduct a two-stage clustering: first, grouping explanations by reasoning intent using GPT-4o; then applying semantic reclustering with \texttt{all-MiniLM-L6-v2} embeddings and Agglomerative Clustering to identify patterns across models. All analyses are fully automated and calibrated on representative samples.

\noindent \textbf{[Results]:} 

\noindent\textbf{\underline{RQ1.1 – Score Distribution and Likert Categorization}}.

\vspace{0.3em}
\noindent\textbf{Global Likert Distribution}.

Table~\ref{tab:likert-global-language}(a) shows that LLMs labeled 65.0\% of all snippets as \emph{readable}, compared to 43.7\% for human annotators and 55.1\% for Scalabrino. Human ratings were more cautious, with 46.5\% falling in the \emph{neutral} category, whereas LLMs and Scalabrino assigned fewer neutral scores (29.4\% and 24.0\%, respectively). Scalabrino also yielded the highest proportion of \emph{unreadable} labels (20.9\%), more than double that of LLMs (5.6\%) and notably higher than humans (9.7\%).

\begin{table}[ht]
\centering
\caption{Likert Distribution of Readability Scores}
\label{tab:likert-global-language}
% === (a) Global Distribution ===
\textbf{(a) Global Distribution}

\scalebox{0.8}{
\begin{tabular}{@{}lccc@{}}
\toprule
\textbf{Source} & \textbf{Neutral (\%)} & \textbf{Readable (\%)} & \textbf{Unreadable (\%)} \\
\midrule
HUMAN       & 46.54 & 43.72 & 9.74 \\
LLM         & 29.35 & 65.03 & 5.62 \\
SCALABRINO  & 23.99 & 55.11 & 20.90 \\
\bottomrule
\end{tabular}
}

% === (b) Distribution by Language ===
\textbf{(b) Distribution by Language}

\scalebox{0.8}{
\begin{tabular}{@{}llccc@{}}
\toprule
\textbf{Lang} & \textbf{Source} & \textbf{Readable (\%)} & \textbf{Neutral (\%)} & \textbf{Unreadable (\%)} \\
\midrule
\multirow{2}{*}{CUDA} 
        & HUMAN & 16.24 & 52.99 & 30.77 \\
        & LLM   & 49.41 & 38.83 & 11.76 \\
\midrule
\multirow{3}{*}{Java} 
        & HUMAN       & 48.89 & 45.48 & 5.63 \\
        & LLM         & 68.08 & 27.46 & 4.46 \\
        & SCALABRINO  & 55.11 & 23.99 & 20.90 \\
\midrule
\multirow{2}{*}{Python} 
        & HUMAN & 55.09 & 43.22 & 1.70 \\
        & LLM   & 70.91 & 25.91 & 3.17 \\
\bottomrule
\end{tabular}
}

\end{table}

\noindent\textbf{Likert Distribution by Language}.
As shown in Table~\ref{tab:likert-global-language}(b), the tendency of LLMs to assign higher readability scores—hereafter referred to as \emph{optimistic bias}—is consistent across languages. For instance, LLMs rated 68.1\% of Java, 70.9\% of Python, and 49.4\% of CUDA snippets as readable, while human annotators rated only 48.9\%, 55.1\%, and 16.2\%, respectively. The gap is especially pronounced for CUDA, where 30.8\% of snippets were marked unreadable by humans—reflecting stricter judgments in more challenging code domains.

\begin{figure}
  \centering
  \includegraphics[width=0.8\textwidth]{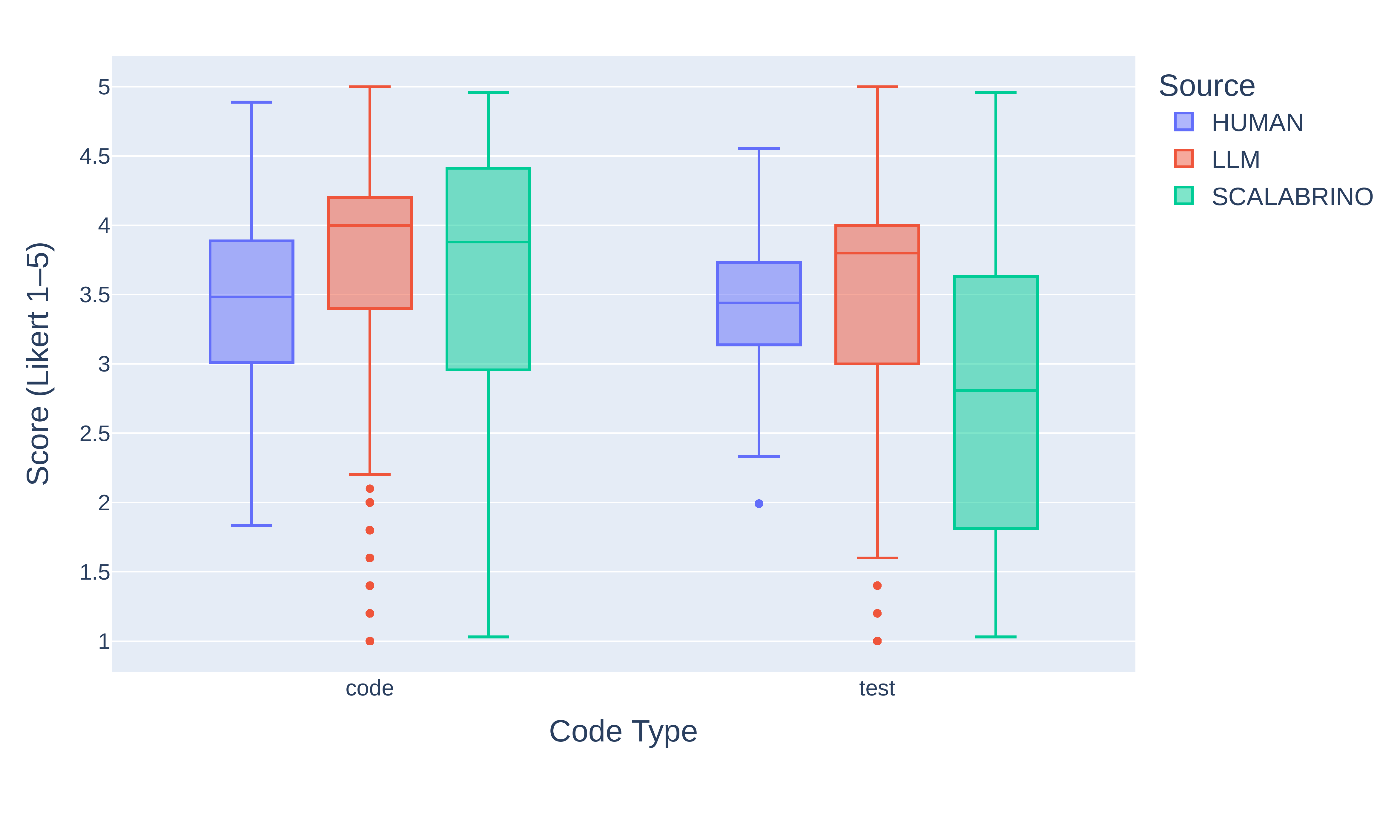} %0.9
  \caption{Likert Distribution by Code Type (Java only).}
    \label{fig:likert_boxplot_by_code_test_java}
\end{figure}

\noindent\textbf{Likert Distribution by Code Type (Java only)}.
Figure~\ref{fig:likert_boxplot_by_code_test_java} contrasts code readability assessments between functional code and unit tests. LLMs rated 66.7\% of Java functional code as readable, versus 53.9\% for unit tests. Human annotators exhibited the opposite trend, labeling 46.1\% of test code as readable, compared to only 43.3\% of functional code. Scalabrino aligned more closely with LLMs on functional code (62.6\% readable), but was substantially harsher on unit tests—assigning only 33.3\% as readable and 40.7\% as unreadable. We hypothesize that this disparity stems from dataset bias. As shown in Table~\ref{tab:datasets-overview}, several datasets (e.g., Buse and Dorn) contain relatively few test snippets, which may hinder Scalabrino’s ability to generalize to test-specific structures. In contrast, LLMs appear more robust across both code types, likely benefiting from broader training data and flexible reasoning capabilities.

\noindent\colorbox{gray!20}{{\parbox{0.98\linewidth}{
\textbf{Finding 1:} LLMs systematically assign higher readability ratings than human annotators across all languages and code types, reflecting a more lenient default rather than true misalignment.
In contrast, humans adopt a more cautious stance, frequently opting for neutral scores. Scalabrino aligns with LLMs on functional code but assigns substantially lower ratings to unit tests—likely due to limited exposure to test-specific patterns during training. These differences indicate divergent evaluation heuristics: statistical learning in LLMs, static rule-based analysis in Scalabrino, and subjective judgment in human evaluations.
}}}

\vspace{0.5em}
\noindent\textbf{\underline{RQ1.2 – Correlation Analysis}}.

\noindent\textbf{Java.}
As shown in Table~\ref{tab:correlation-metrics}, LLMs show moderate alignment with human judgments in Java (Pearson = 0.25, Spearman = 0.25), indicating some consistency in relative scoring, though far from strong agreement. Interestingly, Scalabrino et al.'s model aligns slightly better with human scores (Pearson = 0.31, Spearman = 0.27), consistent with prior findings~\cite{sergeyuk2024reassessing}, suggesting that its predictions—based on lexical, structural, and syntactic features~\cite{scalabrino2018comprehensive}—still capture readability cues valued by developers. In contrast, the correlation between LLMs and Scalabrino remains weak (Pearson = 0.09), indicating divergent scoring strategies.

\begin{table}[ht]
\centering
\caption{Correlation of Readability Scores}
\label{tab:correlation-metrics}
\scalebox{0.8}{
\begin{tabular}{@{}lccc@{}}
\toprule
\textbf{Language} & \textbf{Source Pair} & \textbf{Pearson} & \textbf{Spearman} \\
\midrule
\multirow{1}{*}{CUDA} 
    & LLM vs HUMAN             & 0.00  & -0.01 \\
\midrule
\multirow{3}{*}{Java} 
    & LLM vs HUMAN             & 0.25  & 0.25  \\
    & LLM vs SCALABRINO        & 0.09  & 0.13  \\
    & HUMAN vs SCALABRINO      & 0.31  & 0.27  \\
\midrule
\multirow{1}{*}{Python} 
    & LLM vs HUMAN             & 0.02  & 0.00  \\
\bottomrule
\end{tabular}
}
\end{table}

\noindent\textbf{Python.}
The correlation between LLMs and human scores in Python (Table~\ref{tab:correlation-metrics}) is negligible (Pearson = 0.02, Spearman = 0.003), revealing almost no alignment in ranking. This suggests that LLMs apply different criteria—or inconsistent reasoning—when assessing Python code, potentially due to language-specific stylistic or structural factors not captured in the prompt.

\noindent\textbf{CUDA.}
Similarly, CUDA results (Table~\ref{tab:correlation-metrics}) confirm the lack of alignment: the correlation between LLMs and human scores is essentially null (Pearson = 0.002, Spearman = –0.011). This suggests that LLMs struggle to assess readability consistently in domain-specific or less represented languages. A similar issue was highlighted by Sun et al.~\cite{sun2024source}, who showed that LLMs used as evaluators for source code summarization produced inconsistent judgments across programming languages—particularly for those less frequent in pretraining corpora.

\noindent\colorbox{gray!20}{{\parbox{0.98\linewidth}{
\textbf{Finding 2:}
LLMs show moderate alignment with human readability rankings in Java, but correlations drop significantly in Python and CUDA. While Scalabrino aligns more closely with human scores in Java, no strong baseline exists for other languages. These results suggest that LLMs may struggle to assess readability reliably outside of well-represented training domains, underscoring the need for improved cross-language robustness.
}}}

%%%%%%%%%%%%%%%%%%%%%%%%%%%%%%%%%%%%%%%%%%%%%%

\noindent\textbf{\underline{RQ1.3 – Sentiment Polarity of Justifications}}.

\begin{figure}[ht]
\centering
\scalebox{1.02}{
\begin{minipage}{0.49\linewidth}
    \centering
    \includegraphics[width=\linewidth]{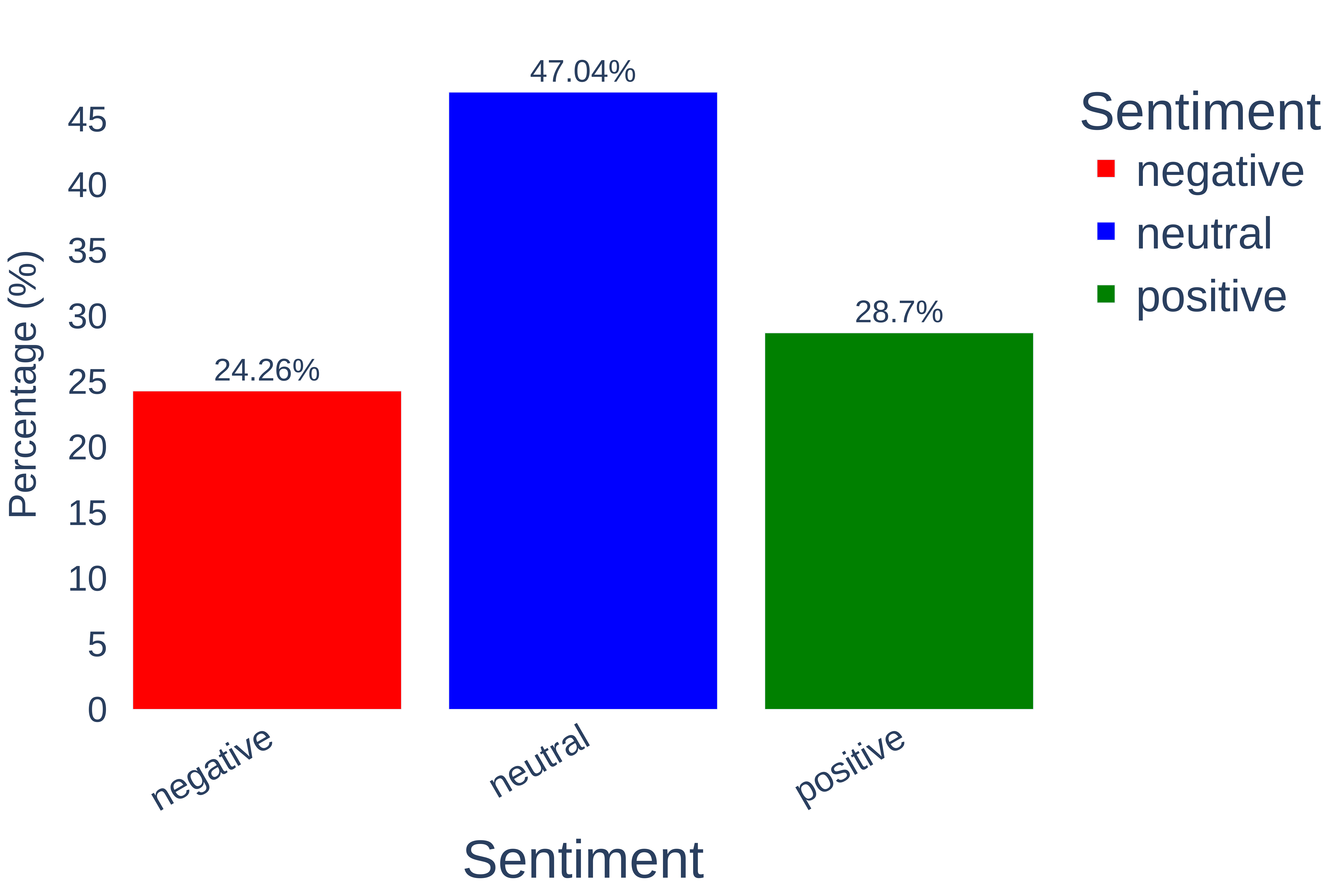}
    \vspace{-0.5em}
    
    \textbf{(a) GPT-4o}
\end{minipage}
\hfill
\begin{minipage}{0.49\linewidth}
    \centering
    \includegraphics[width=\linewidth]{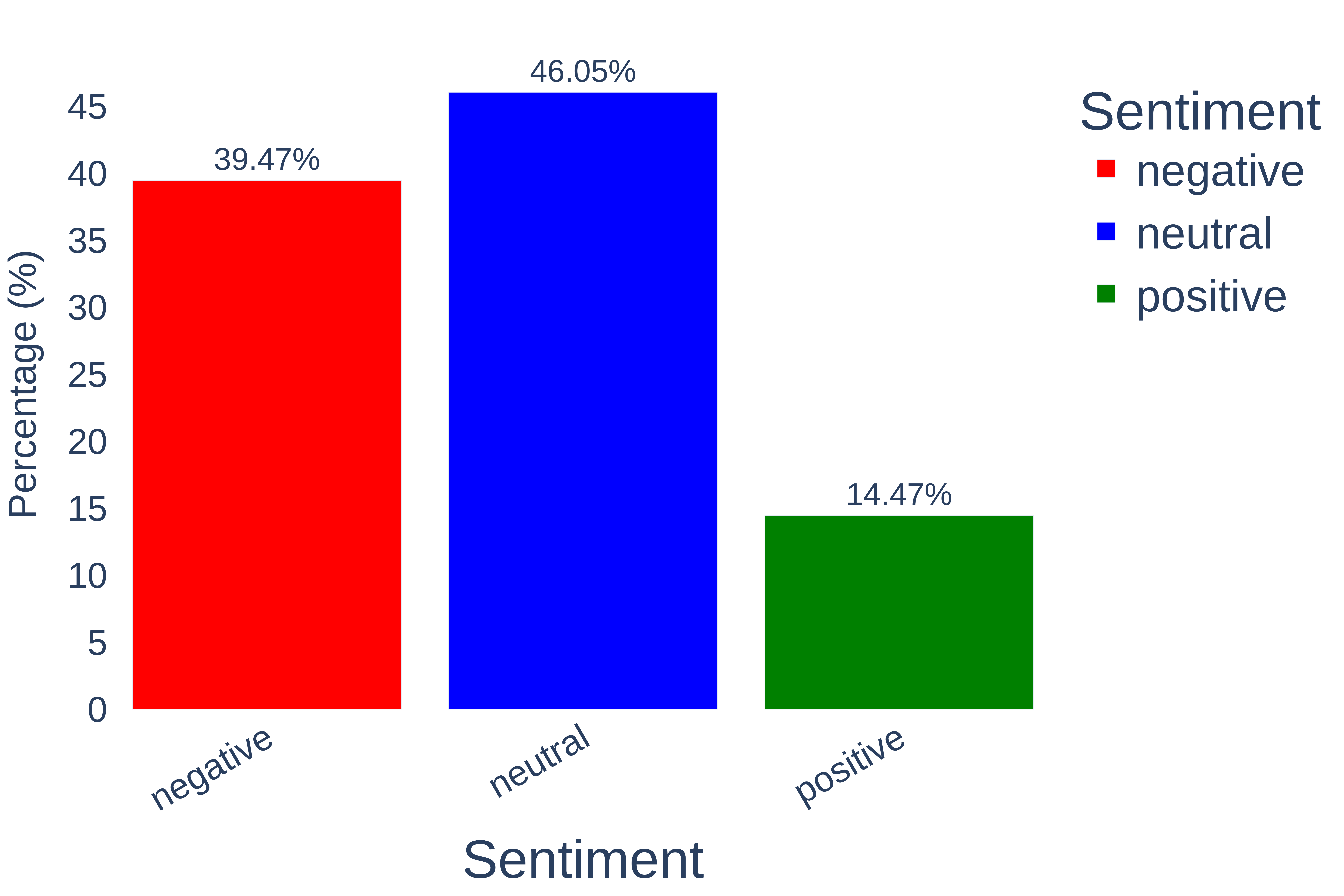}
    \vspace{-0.5em}
    
    \textbf{(b) RoBERTa-TweetEval}
\end{minipage}
}
\vspace{0.5em}
\caption{Sentiment Polarity of Justifications by GPT-4o and RoBERTa-TweetEval.}
\vspace{0.3em}
\label{fig:sentiment_polarity}
\end{figure}

\noindent\textbf{GPT-4o Analysis.}
The large-scale evaluation with GPT-4o reveals a relatively balanced distribution: \textbf{47.8\%} of rationales were labeled as \textit{neutral}, \textbf{25.9\%} as \textit{positive}, and \textbf{26.3\%} as \textit{negative}. This suggests that LLMs tend to adopt an analytical and descriptive tone when explaining readability, while still offering both praise and criticism.

\noindent\textbf{RoBERTa-TweetEval Classifier.}
The local classifier produced a slightly more polarized distribution: \textbf{46.1\%} neutral, \textbf{14.5\%} positive, and \textbf{39.5\%} negative. The higher share of negative sentiment highlights a tendency to identify and emphasize flaws in the code, such as poor naming or lack of clarity, even when the overall score is not low.

\noindent\textbf{Interpretation.}
The consistency across both methods—particularly in the prevalence of neutral-toned responses—suggests that LLMs generally provide balanced and descriptive explanations rather than overly favorable ones. Still, the higher proportion of negative sentiment detected by the local classifier indicates that LLMs often identify specific issues (e.g., poor naming, lack of clarity) even when overall readability scores are moderate. This underscores the value of jointly analyzing both numerical ratings and justification tone to better understand LLM-based assessments.

\vspace{0.5em}
\noindent\colorbox{gray!20}{{\parbox{0.98\linewidth}{
\textbf{Finding 3:}
LLM justifications tend to adopt a neutral tone, with both positive and negative sentiments appearing in similar proportions. This balance—observed across GPT-4o and RoBERTa analyses—suggests that LLMs avoid overly enthusiastic or critical language, instead offering descriptive feedback that focuses on specific code characteristics. Such tone consistency supports the interpretability of LLM rationales.
}}}

%%%%%%%%%%%%%%%%%%%%%%%%%%%%%%%%%%%%%%%%%%%%
\vspace{0.5em}
\textbf{RQ1.4 – Aspect Detection and Reasoning Clusters}
\noindent\textbf{Aspect Frequency Analysis.}

\begin{figure}[ht]
\centering
\scalebox{0.6}{
\begin{minipage}{\linewidth}
    \centering
    \includegraphics[width=\linewidth]{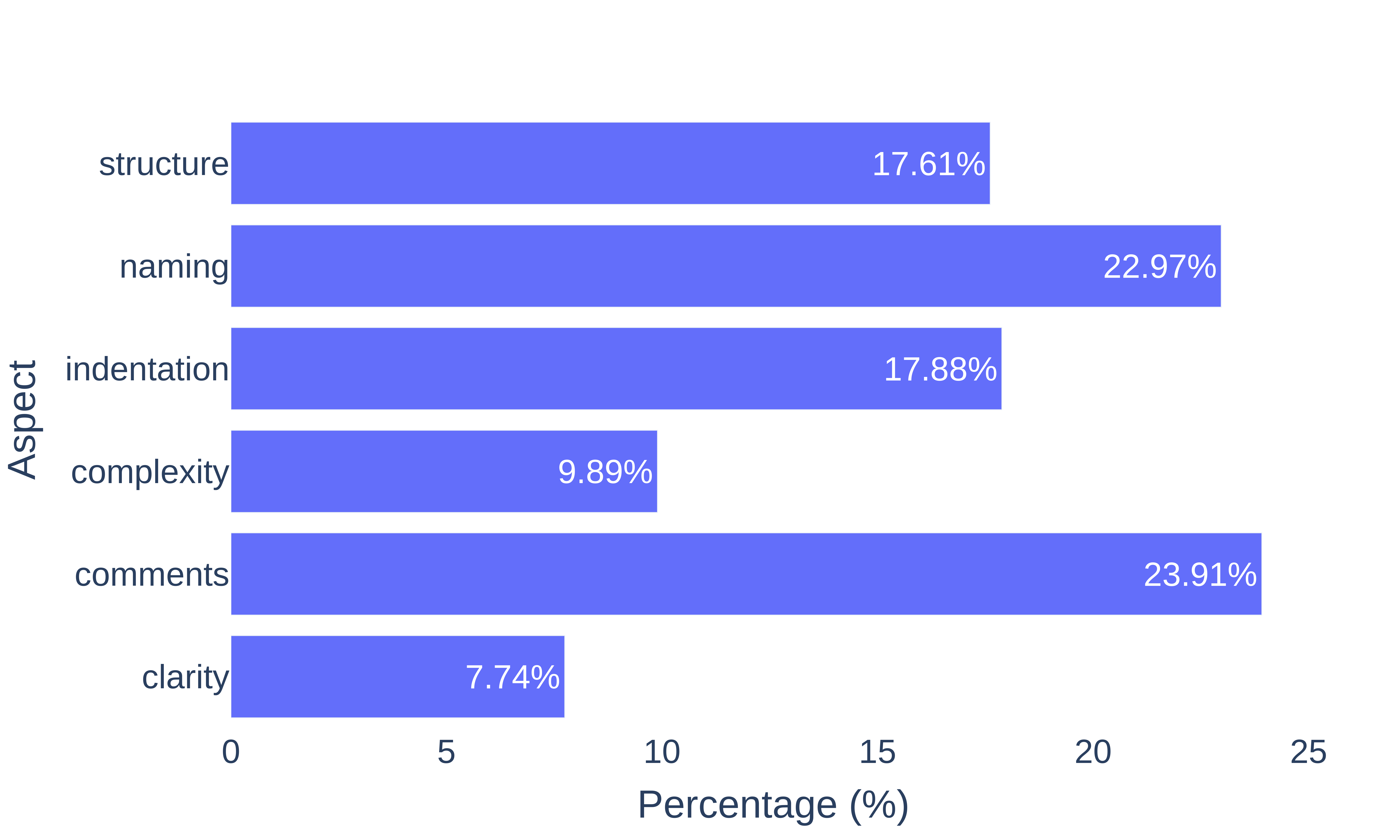}
    \textbf{(a) Aspect Frequency Analysis}
    \phantomsection
    \label{fig:aspect-distribution}
    \includegraphics[width=\linewidth]{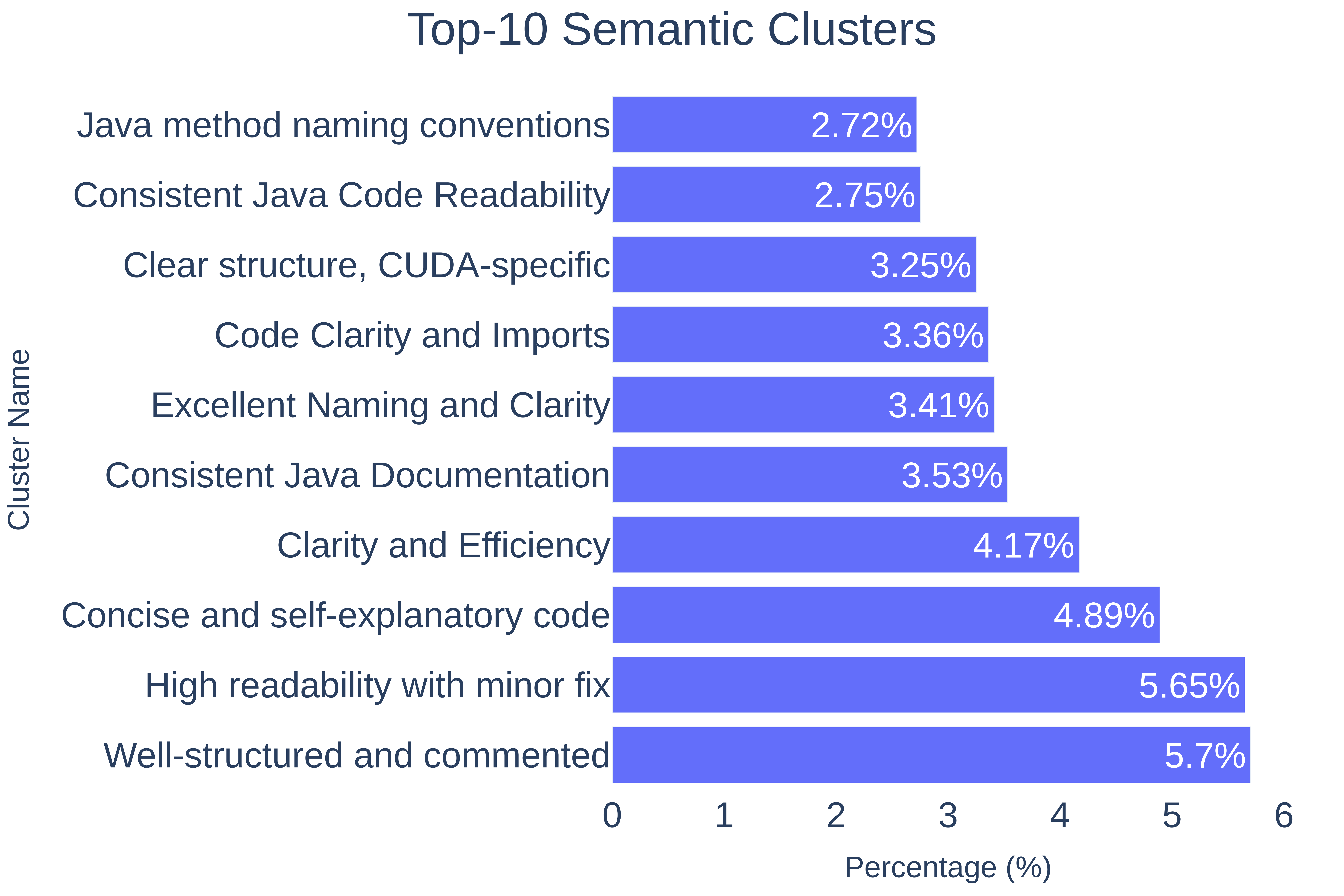}
    \textbf{(b) Reasoning Clusters Analysis}
    \phantomsection
    \label{fig:cluster-distribution}
\end{minipage}
}

\caption{Aspect mentions and reasoning structures in LLM justifications.}
\label{fig:aspect_andreasoning}
\end{figure}

The analysis of aspect occurrences in LLM-generated justifications (Figure~\ref{fig:aspect_andreasoning}(a)) shows that three main dimensions dominate: comments (23.91\%), naming (22.97\%), and indentation/structure (17.88\% / 17.61\%). Together, these four cover more than 82\% of all references, indicating that LLMs predominantly focus on surface-level code features and formatting conventions when assessing readability. Surprisingly, higher-order concerns such as clarity (7.74\%) and complexity (9.89\%) are mentioned less frequently, suggesting that LLMs tend to emphasize syntactic polish over semantic depth in their judgments.

This surface-level bias echoes prior observations in the code readability literature~\cite{sergeyuk2024reassessing}, where tools and models often prioritize formatting cues due to their easier detectability and lower ambiguity. Yet, in educational or professional settings, deeper issues like logical clarity and cognitive load are often more decisive in human assessments—highlighting a potential gap between LLM reasoning and human expectations.

\noindent\textbf{Reasoning Clusters Analysis.}

The initial GPT-4o explanations yielded 11,020 unique reasoning labels, reflecting highly diverse phrasing and articulation. To reduce redundancy, we applied semantic reclustering using all-MiniLM-L6-v2 and Agglomerative Clustering, yielding 100 canonical clusters. The distribution (Figure~\ref{fig:aspect_andreasoning}(b) remains long-tailed: the top cluster—\textit{“Well-structured and commented”}—covers \textbf{5.70\%} of all instances, followed by \textit{“High readability with minor fix”} (5.65\%), and \textit{“Concise and self-explanatory code”} (4.89\%). This high variability suggests nuanced but inconsistent reasoning behaviors, in contrast to human rubrics which typically rely on fewer, well-defined evaluation criteria.

\vspace{0.5em}
\noindent\colorbox{gray!20}{{\parbox{0.98\linewidth}{
\textbf{Finding 4:}
LLM justifications primarily focus on surface-level features such as comments, naming, and indentation. Yet, the overall reasoning space is highly fragmented: GPT-4o assigned over 11,000 distinct labels across all LLMs. To mitigate redundancy, we applied semantic reclustering using all-MiniLM-L6-v2 embeddings and agglomerative clustering. Despite this, the large number of resulting clusters reveals a lack of convergence in how LLMs explain readability, underscoring the need for more structured and consistent prompting strategies.
}}}

%%%%%%%%%%%%%%%%%%%%%%%%%%%%%%%%%%%%%%%%%%%%%%%%%%%%%%%%%%%%%%%
\vspace{0.5em}
Our findings on LLM-based readability assessment align with and expand upon recent studies that question the robustness and consistency of LLMs as code evaluators. Wang et al.~\cite{wang2025can} show that while LLMs can approximate human preferences in general software engineering tasks, their behavior varies significantly across languages, code domains, and prompts. Specifically, they find that LLMs often overfit to superficial cues and lack robustness in less common domains—mirroring our observation of weak correlation in Python and CUDA. Furthermore, Vitale et al.~\cite{vitale2025personalized} show that current LLMs, even when personalized via few-shot learning fail to surpass feature-based models like Scalabrino et al~\cite{scalabrino2018comprehensive} in predicting human readability judgments. This reinforces our conclusion that LLMs lack an embedded, stable representation of "readability" and rely on shallow heuristics. Interestingly, both works converge with our sentiment and aspect-based analyses: LLMs tend to focus on easily observable features such as formatting and naming, rather than semantic clarity or structural complexity. Vitale et al.~\cite{vitale2025personalized} further highlight that readability judgments by humans are often self-contradictory, with 20–30\% of assessments being inconsistent even for the same developer—a challenge we mitigate by using average labels and clustering. Altogether, these results emphasize that LLMs, despite producing plausible explanations, still lack principled and generalizable heuristics for assessing code readability. Progress will likely require improved datasets, enhanced grounding in semantic aspects, and more explicit guidance via human-aligned prompts.

\vspace{0.5em}
\noindent\highlight{Summary of \textbf{RQ1:} LLMs systematically assign higher readability scores than both humans and static models, revealing an optimistic bias. While their justifications are generally balanced in tone, they predominantly focus on surface-level features like naming and indentation. Alignment with human scores is moderate in Java, but negligible in Python and CUDA, highlighting the limitations of current LLMs in cross-language generalization. Despite the richness of their rationales, LLMs show fragmented reasoning patterns, with high variance in explanation styles and limited semantic depth.}

%%%%%%%%%%%%%%%%%%%%%%%%%%%%%%%%%%%%%%%%%%%%%%%%%%%%%%%%%%%%%%%%%
\subsection{RQ2: Impact of LLM configuration on evaluation behavior}
\label{subsec:RQ2}
\noindent \textbf{[Experimental design]:} 
This research question investigates how different configuration factors affect the behavior of LLMs in code readability evaluation, focusing on both the assigned scores and the reasoning quality expressed in the justifications. We decompose RQ2 into four sub-questions. First, \textbf{RQ2.1 (persona framing)} examines whether the use of different developer personas (junior versus senior) has a measurable impact on scoring behavior. We compare the mean scores, standard deviations, and mean absolute errors (MAE) relative to human judgments across persona conditions. Second, \textbf{RQ2.2 (prompting strategy)} evaluates the effect of different prompting paradigms namely Zero-Shot (ZSL), Few-Shot (FSL), Chain-of-Thought (CoT), and Tree-of-Thought (ToT) on evaluation accuracy and variability. Each strategy is applied uniformly across models and code types to isolate its influence on model responses. Third, \textbf{RQ2.3 (decoding parameters)} assesses how randomness in generation affects score stability. We systematically vary temperature and top\_p settings, and compute their correlations with score variability using both Pearson’s $r$ and Spearman’s $\rho$. Finally, \textbf{RQ2.4 (model scale)} analyzes differences between small-scale ($\leqslant7B$ parameters) and large-scale ($\geqslant64B$ parameters) models. Here, we focus on the extent to which model capacity influences output consistency and alignment with human ratings. Together, these four sub-questions provide a comprehensive view of how LLM configuration choices shape their behavior as code readability evaluators.

\noindent \textbf{[Results]:} 

%\vspace{0.5em}
\noindent\textbf{RQ2.1 — Persona (Junior vs. Senior)}.
The results in Table~\ref{tab:llm-configuration-comparison}(a) show minimal differences between the two assigned personas. Senior models achieve a slightly higher average score (3.77 vs. 3.75), while junior models yield a marginally lower mean absolute error (0.69 vs. 0.70). Standard deviations are identical (0.68 for scores, 0.45 for MAE), indicating near-identical evaluation behavior. 

\vspace{0.5em}
\noindent\colorbox{gray!20}{{\parbox{0.98\linewidth}{
\textbf{Finding 5:}
Persona assignment (Junior vs. Senior) has minimal effect on code readability evaluation outcomes. Prompt structure appears to dominate model behavior, overriding subtle persona-driven nuances.
}}}

\begin{table}[ht]
\centering
\caption{Impact of LLM Configuration on Readability Scoring Accuracy and Consistency}
\label{tab:llm-configuration-comparison}
\scalebox{0.7}{
\begin{tabular}{@{}c@{}}

% === Persona ===
\begin{minipage}[t]{\linewidth}
\centering
\textbf{(a) Persona Comparison: Junior vs. Senior Profiles}
\label{tab:rq2a-persona}
% \vspace{0.5em}
\begin{tabular}{lcccc}
\toprule
\textbf{Persona} & \textbf{Score Mean} & \textbf{Score Std} & \textbf{MAE Mean} & \textbf{MAE Std} \\
\midrule
Senior & 3.77 & 0.68 & 0.70 & 0.45 \\
Junior & 3.75 & 0.68 & 0.69 & 0.45 \\
\bottomrule
\end{tabular}
\end{minipage}
\\%[0.8em]
% \vspace{0.5em}
% === Prompting ===
\begin{minipage}[t]{\linewidth}
\centering
\textbf{(b) Prompting Strategy Comparison}
\label{tab:rq2b-prompting}

\begin{tabular}{lcccc}
\toprule
\textbf{Prompt} & \textbf{Score Mean} & \textbf{Score Std} & \textbf{MAE Mean} & \textbf{MAE Std} \\
\midrule
ZSL & 3.57 & 0.71 & 0.89 & 0.63 \\
FSL & 3.86 & 0.72 & 1.08 & 0.69 \\
CoT & 3.70 & 0.65 & 0.93 & 0.64 \\
ToT & 3.92 & 0.55 & 1.04 & 0.67 \\
\bottomrule
\end{tabular}
\end{minipage}
\\[0.8em]
% === Decoding ===
\begin{minipage}[t]{\linewidth}
\centering
\textbf{(c) Decoding Parameter Correlation}
\label{tab:rq2c-decoding}

\begin{tabular}{lcccc}
\toprule
\textbf{Parameter} & \textbf{Pearson $r$} & \textbf{$p$-value} & \textbf{Spearman $\rho$} & \textbf{$p$-value} \\
\midrule
Temperature & -0.014 & $2.7\times10^{-138}$ & -0.009 & $7.3\times10^{-56}$ \\
Top\_p     & -0.011 & $7.5\times10^{-87}$ & -0.007 & $8.6\times10^{-37}$ \\
\bottomrule
\end{tabular}
\end{minipage}
\\[0.8em]
% === Models ===
\begin{minipage}[t]{\linewidth}
\centering
% \vspace{0.5em}
\textbf{(d) Model Scale Comparison}
\label{tab:rq2d-models}

\begin{tabular}{lcccc}
\toprule
\textbf{Model} & \textbf{Score Mean} & \textbf{Score Std} & \textbf{MAE Mean} & \textbf{MAE Std} \\
\midrule
Gemma 7B IT                         & 4.01 & 0.48 & 0.75 & 0.44 \\
Falcon 3 7B                         & 3.99 & 0.45 & 0.73 & 0.41 \\
Deepseek LLM 7B Chat                & 3.96 & 0.74 & 0.86 & 0.51 \\
Phi-3 Mini                          & 3.90 & 0.58 & 0.72 & 0.46 \\
Mistral 7B Instruct (v0.3)          & 3.80 & 0.67 & 0.72 & 0.47 \\
Mistral Large (v2407)               & 3.68 & 0.54 & 0.56 & 0.41 \\
Deepseek Chat                       & 3.67 & 0.62 & 0.62 & 0.42 \\
Claude 3 Sonnet                     & 3.65 & 0.73 & 0.67 & 0.45 \\
GPT-4o                              & 3.63 & 0.77 & 0.71 & 0.48 \\
Qwen2.5 7B Instruct                 & 3.39 & 0.81 & 0.72 & 0.45 \\
\bottomrule
\end{tabular}
\end{minipage}
\\
\end{tabular}
}
\end{table}

\vspace{0.5em}
\noindent\textbf{RQ2.2 — Prompting Strategy}.
Table~\ref{tab:llm-configuration-comparison}(b) compares the effect of prompting strategies on readability evaluation. ToT (Tree-of-Thought) prompts yield the highest average scores (3.92), followed by FSL (3.86), CoT (3.70), and ZSL (3.57). However, this increase in score does not consistently translate to improved alignment with human ratings: FSL and ToT both produce the highest mean absolute errors (1.08 and 1.04), while ZSL achieves the lowest MAE (0.89).

\vspace{0.5em}
\noindent\colorbox{gray!20}{{\parbox{0.98\linewidth}{
\textbf{Finding 6:}
Elaborate prompting styles (e.g., ToT, FSL) yield higher scores but deviate more from human ratings. Simpler prompts like ZSL better align with human judgment, highlighting a trade-off between expressiveness and accuracy.
}}}

\vspace{0.5em}
\noindent\textbf{RQ2.3 — Decoding Parameters.}
The correlation analysis (Table~\ref{tab:llm-configuration-comparison}(c)) reveals a very weak but statistically significant negative relationship between decoding parameters and LLM-assigned readability scores. Specifically, Pearson’s $r$ is $-0.014$ ($p < 10^{-130}$) and Spearman’s $\rho$ is $-0.009$ ($p < 10^{-50}$) for \texttt{temperature}, while \texttt{top-p} shows even lower correlation magnitudes ($r = -0.011$, $\rho = -0.007$). Although the effect sizes are negligible, the large sample size makes the findings statistically robust. These results suggest that higher generation randomness slightly decreases the consistency of readability assessments. This aligns with Sun et al.~\cite{sun2024source}, who found that higher \texttt{temperature} reduces coherence, while \texttt{top-p} has a milder impact. They recommend lower values for more stable outputs.

\vspace{0.5em}
\noindent\colorbox{gray!20}{{\parbox{0.98\linewidth}{
\textbf{Finding 7:}
Code readability scores are largely unaffected by decoding parameters, with only negligible but statistically significant negative correlations observed. This suggests that generation randomness (\texttt{temperature}, \texttt{top\_p}) plays a minimal role in shaping LLM-based code readability assessments.
}
}}

\vspace{0.5em}
\noindent \textbf{RQ2.4 — Model Family and Size.}
We observe systematic differences across LLMs based on their parameter size (Table~\ref{tab:llm-configuration-comparison}(d)). Small-scale models such as \texttt{Gemma 7B} (4.01), \texttt{Falcon 3 7B} (3.99), and \texttt{Deepseek LLM 7B Chat} (3.96) tend to assign higher readability scores, with MAE values ranging from 0.72 to 0.86. In contrast, large-scale models such as \texttt{GPT-4o} (3.63), \texttt{Claude 3 Sonnet} (3.65), and \texttt{Mistral Large v2407} (3.68) yield lower scores but generally achieve better alignment with human judgments—especially \texttt{Mistral Large} (MAE = 0.56) and \texttt{Deepseek Chat} (MAE = 0.62).

\vspace{0.5em}
\noindent\colorbox{gray!20}{{\parbox{0.98\linewidth}{
\textbf{Finding 8:} Small-scale models assign higher readability scores but align less consistently with human annotations, whereas large-scale models yield lower scores with improved agreement. Model scale plays a key role in shaping both scoring behavior and alignment fidelity.
}}}

%%%%%%%%%%%%%%%%%%%%%%%%%%%%%%%%%%%%%%%%%%%%%%%%%%%%%%%%%%%%%%%%%%%%%%%%%%%%%%%%%%%%%%%%%%%%%%%%%%%%%%%%%%%%%%
\vspace{0.5em}
Our results highlight that LLM evaluation behavior is shaped more by prompt structure and model scale than by decoding parameters or superficial persona framing. This is consistent with findings by Vitale et al.~\cite{vitale2025personalized}, who show that prompting strategies have a more significant influence than personalization efforts in readability prediction. Similarly, in code summarization tasks, Sun et al.~\cite{sun2024source} observe that Chain-of-Thought (CoT) and few-shot prompting may introduce verbosity and over-rationalization, leading to divergence from human-style outputs—mirroring our observation that elaborate prompting (FSL, ToT) increases scores but reduces alignment with human judgments. The impact of decoding parameters (temperature, top-$p$) is negligible in our setup, confirming prior work~\cite{sun2024source} which recommends low randomness for stability in source code tasks. Finally, our observation that large-scale models align better with human readability ratings resonates with Wang et al.~\cite{wang2025can}, who find that output-based methods relying on larger LLMs (e.g., GPT-4o, DeepSeek) achieve near-human evaluation performance in code-related SE tasks. These results collectively suggest that \textbf{scale and structure}, rather than superficial variations like persona or decoding, are the most critical levers for building human-aligned readability evaluators.

\vspace{0.5em}
\noindent\highlight{Summary of \textbf{RQ2:} Prompt structure and model scale have the strongest influence on LLM readability evaluation behavior. Persona framing has negligible effect, and decoding parameters like \texttt{temperature} or \texttt{top\_p} only introduce minor score variance. Complex prompting strategies (e.g., CoT, FSL, ToT) increase the verbosity and optimism of model outputs, but often reduce alignment with human judgments. Large-scale models exhibit lower score inflation and improved correlation with human labels, making them better suited for readability evaluation in SE contexts.}

%%%%%%%%%%%%%%%%%%%%%%%%%%%%%%%%%%%%%%%%%%%%%%%%%%%%%%%%%%%%%%%%%%%%%%%%%%%%%%%%%%%%%%%%%%%%%%%%%%%%%%%%%%%%%%%
\subsection{RQ3: Impact of human-centric guidance on LLM evaluation accuracy}
\label{subsec:RQ3}

\noindent \textbf{[Experimental design]:} 
To assess the impact of developer-guided prompting on the quality of LLM-based readability evaluation, we compare two prompt configurations: a \textbf{generic baseline} using standard criteria (e.g., naming clarity, structure, indentation, comments, complexity, and overall clarity), and a \textbf{developer-guided} variant embedding top six human-centric criteria tailored to the code type and developer persona. Both configurations are evaluated on 656 code snippets across three languages (Java, Python, CUDA) and two categories (functional and test code). 
The generic baseline aggregates results across all four prompting strategies (ZSL, FSL, CoT, ToT) applied without developer-specific guidance. This aggregated baseline is compared against the corresponding aggregation under developer-guided prompting to assess score alignment and consistency. 
We evaluate outcomes along three main dimensions. First, we measure \textbf{alignment with human judgments} using the Mean Absolute Error (MAE) between model predictions and human-annotated readability scores. Second, we assess \textbf{score stability} by computing the standard deviation of LLM-generated scores across models for each code snippet. Third, we examine \textbf{reasoning quality} by analyzing the coverage of readability aspects in LLM justifications. To contextualize LLM performance, we include the static model of Scalabrino et al.~\cite{scalabrino2018comprehensive} as a baseline for Java code, comparing its MAE and score variance against those of generic and guided LLM configurations. Statistical significance is assessed using one-sided Mann–Whitney U tests on per-snippet MAE distributions to ensure robust non-parametric comparison.

\noindent \textbf{[Results]:}

\noindent \textbf{RQ3.1 — Score Alignment}.
Table~\ref{tab:mae-metrics} summarizes the Mean Absolute Error (MAE) across prompt configurations and code subsets, including Scalabrino et al.’s static model~\cite{scalabrino2018comprehensive} as a baseline for Java. Overall, the global MAE remains stable between generic (0.983) and developer-guided (0.996) prompts, indicating that explicit guidance does not systematically improve alignment at scale. At the language level, results show distinct behaviors depending on code type. For \textbf{Java production code}, the developer-guided configuration slightly increases the MAE (1.01) relative to the generic baseline (0.97), while still achieving performance comparable to Scalabrino’s model (1.09). In contrast, for \textbf{Java test code}, both LLM configurations substantially outperform Scalabrino (1.19), with the developer-guided version achieving the lowest MAE (0.83 vs.\ 0.88 for the generic baseline). These findings suggest that while structured prompts do not yield uniform gains across all contexts, they significantly enhance alignment in code domains characterized by clearer structural conventions, such as test methods.

\begin{table}[ht]
\centering
\vspace{0.3em}
\caption{MAE for Generic vs. Human-Centric Prompts}
\label{tab:mae-metrics}
\vspace{0.3em}
\scalebox{0.8}{
\begin{tabular}{l|ccc|ccc}
\toprule
\multirow{2}{*}{\textbf{Subset}} & \multicolumn{3}{c|}{\textbf{MAE Mean}} & \multicolumn{3}{c}{\textbf{MAE Std}} \\
 & Generic & Guided & Scalabrino & Generic & Guided & Scalabrino \\
\midrule
Global      & 0.983 & 0.996 & -- & 0.658 & 0.669 & -- \\
Java – Code & 0.967 & 1.012 & 1.091 & 0.638 & 0.661 & 0.698 \\
Java – Test & 0.882 & 0.835 & 1.193 & 0.653 & 0.625 & 0.771 \\
\bottomrule
\end{tabular}
}
\end{table}

\vspace{0.5em}
\noindent\colorbox{gray!20}{{\parbox{0.98\linewidth}{
\textbf{Finding 9:}  
Developer-guided prompting does not consistently improve LLM alignment at a global level, yet it enhances evaluation accuracy in structured settings such as Java test code. On production code, guided LLMs perform comparably to Scalabrino’s static model, whereas on test code, they clearly surpass it. These results indicate that explicit, aspect-based prompting helps LLMs better capture readability dimensions associated with standardized structures and testing practices.
}}}

\vspace{0.5em}
\noindent \textbf{ RQ3.2 — Score Stability}.
We analyze the variability of readability scores produced by different evaluation strategies. Table~\ref{tab:readability-score-comparison} reports the mean readability scores and their per-snippet standard deviations across evaluators. At the global level, developer-guided prompts yield a slightly lower average score than the generic baseline (0.59 vs.\ 0.60), but exhibit substantially higher variance (0.22 vs.\ 0.13), indicating reduced scoring stability. This trend is particularly pronounced in \textbf{Java test code}, where standard deviation rises from 0.12 (generic) to 0.25 (guided).  
Scalabrino’s static model produces the highest mean scores (e.g., 0.85 on tests) and the highest variance across all subsets, with standard deviations reaching 0.60 (Java code) and 0.57 (Java test). These values suggest a broader spread in its predictions, possibly due to rigid feature-based heuristics that amplify sensitivity to local code variations.

\begin{table}[ht]
\centering
\vspace{0.3em}
\caption{Mean and Variance of Readability Scores by Evaluator and Subset}
\label{tab:readability-score-comparison}
\vspace{0.3em}
\scalebox{0.72}{
\begin{tabular}{l|ccc|ccc}
\toprule
\multirow{2}{*}{\textbf{Subset}} & \multicolumn{3}{c|}{\textbf{Mean Score}} & \multicolumn{3}{c}{\textbf{Std Dev}} \\
                                 & Generic & Guided & Scalabrino           & Generic & Guided & Scalabrino \\
\midrule
Global                          & 0.60    & 0.59   & --                & 0.13    & 0.22   & -- \\
Java – Code                     & 0.54    & 0.46   & 0.77                 & 0.13    & 0.18   & 0.60 \\
Java – Test                     & 0.56    & 0.33   & 0.85                 & 0.12    & 0.25   & 0.57 \\
\bottomrule
\end{tabular}
}
\end{table}

Mann–Whitney U tests (Table~\ref{tab:mannwhitney-score-variance}) confirm that the increase in score variability between generic and guided LLM prompting is statistically significant across all subsets ($p < 0.001$). However, comparisons involving Scalabrino do not reveal statistically significant differences in variance ($p > 0.99$), suggesting that although LLMs and static models differ in scoring behavior, their overall dispersion patterns may be similarly wide.
These results highlight a trade-off introduced by developer-guided prompting: while guidance may improve interpretability and alignment in structured contexts (cf.\ RQ3.1), it can also promote divergence in scoring behavior across models, potentially reflecting increased sensitivity to nuanced criteria or a lack of consensus on how to weigh them.

\begin{table}[ht]
\centering
\caption{Mann–Whitney U Test on Score Variability}
%\vspace{0.3em}
\label{tab:mannwhitney-score-variance}
\scalebox{0.8}{
\begin{tabular}{l|l|l|r|r}
\toprule
\textbf{Subset} & \textbf{Group 1} & \textbf{Group 2} & \textbf{U Statistic} & \textbf{p-value} \\
\midrule
Global      & Generic     & Guided      & 664{,}036.0 & 0.000742 \\
Java – Code & Generic     & Guided      & 649{,}944.0 & 0.000000 \\
Java – Test & Generic     & Guided      & 203{,}571.0 & 0.000000 \\
Global      & Generic     & Scalabrino  & 94{,}240.0  & 0.999667 \\
Global      & Guided      & Scalabrino  & 90{,}115.0  & 0.999989 \\
\bottomrule
\end{tabular}
}
\end{table}

\vspace{0.5em}
\noindent\colorbox{gray!20}{{\parbox{0.98\linewidth}{
\textbf{Finding 10:}  
Developer-guided prompting significantly increases score variance among LLMs, especially for Java test code. This suggests that explicit criteria amplify interpretive diversity across models, reducing scoring stability. Although Scalabrino’s static model exhibits even greater variance, differences with LLMs are not statistically significant, indicating that both approaches may struggle with consistency in nuanced evaluation scenarios.
}}}

\vspace{0.5em}
%%%%%%%%%%%%%%%%%%%%%%%%%%%%%%%%%%%
\noindent \textbf{RQ3.3 — Aspect Coverage}.
Across all languages and code types, developer-guided prompts substantially increase aspect coverage (Table~\ref{tab:aspect-coverage}). In Java production code (Table~\ref{tab:aspect-java-code}(a)), mentions of core aspects like Structure (92.23\% vs. 72.08\%), Nesting (84.80\% vs. 12.00\%), and Goal (70.25\% vs. 7.00\%) show dramatic gains. A similar trend appears in Java test code (Table~\ref{tab:aspect-java-test}(b)), where Assertions rise from 15.00\% to 94.38\% and Purpose Clarity from 9.00\% to 76.63\%. The improvement is also pronounced in Python (Table~\ref{tab:aspect-python-code}(c)), with Structure rising from 18.00\% to 90.00\%, and in CUDA (Table~\ref{tab:aspect-cuda-code}(d)), where Reading Flow improves from 10.00\% to 68.21\%. These consistent gains indicate that prompts grounded in developer language strongly steer LLMs toward justifications that reflect human reasoning patterns.

\begin{table}[ht]
\centering
\vspace{0.3em}
\caption{Aspect Coverage (\%): Generic vs. Human-Centric Prompts}
\label{tab:aspect-coverage}

\vspace{0.5em}
% Ligne 1
\begin{minipage}{0.48\linewidth}
\centering
\textbf{(a) Java – Code}\label{tab:aspect-java-code}
\vspace{0.3em}

\scalebox{0.72}{
\begin{tabular}{lrr}
\toprule
\textbf{Aspect} & Gen. & Hum. \\
\midrule
Structure & 72.08 & 92.23 \\
Nesting & 12.00 & 84.80 \\
Goal & 7.00 & 70.25 \\
Length & 10.50 & 73.52 \\
Inline & 6.00 & 48.48 \\
Flow & 12.50 & 74.98 \\
\bottomrule
\end{tabular}
}
\end{minipage}
\hfill
\begin{minipage}{0.48\linewidth}
\centering
\textbf{(b) Java – Test}\label{tab:aspect-java-test}
\vspace{0.3em}

\scalebox{0.72}{
\begin{tabular}{lrr}
\toprule
\textbf{Aspect} & Gen. & Hum. \\
\midrule
Structure & 75.49 & 92.90 \\
Names & 92.39 & 94.38 \\
Assert. & 15.00 & 94.38 \\
Size & 12.00 & 68.64 \\
Purpose & 9.00 & 76.63 \\
Fixture & 6.00 & 61.54 \\
\bottomrule
\end{tabular}
}
\end{minipage}

\vspace{1em}

% Ligne 2
\begin{minipage}{0.48\linewidth}
\centering
\textbf{(c) Python – Code}\label{tab:aspect-python-code}
\vspace{0.3em}

\scalebox{0.72}{
\begin{tabular}{lrr}
\toprule
\textbf{Aspect} & Gen. & Hum. \\
\midrule
Structure & 18.00 & 90.00 \\
Nesting & 5.00 & 81.00 \\
Goal & 6.00 & 72.00 \\
Length & 4.00 & 67.00 \\
Inline & 3.00 & 45.00 \\
Flow & 8.00 & 70.00 \\
\bottomrule
\end{tabular}
}
\end{minipage}
\hfill
\begin{minipage}{0.48\linewidth}
\centering
\textbf{(d) CUDA – Code}\label{tab:aspect-cuda-code}
\vspace{0.3em}

\scalebox{0.72}{
\begin{tabular}{lrr}
\toprule
\textbf{Aspect} & Gen. & Hum. \\
\midrule
Structure & 64.71 & 84.89 \\
Nesting & 10.00 & 71.16 \\
Goal & 6.50 & 48.95 \\
Length & 9.00 & 54.57 \\
Inline & 5.00 & 48.69 \\
Flow & 10.00 & 68.21 \\
\bottomrule
\end{tabular}
}
\end{minipage}
\vspace{0.5em}
\end{table}

\vspace{0.5em}
\noindent\colorbox{gray!20}{{\parbox{0.98\linewidth}{
\textbf{Finding 11:}
Developer-guided prompts drastically improve the coverage of developer-endorsed readability aspects in LLM justifications, enhancing interpretability and alignment with human evaluation criteria.
}}}

%%%%%%%%%%%%%%%%%%%%%%%%%%%%%%%%%%%%%
Our findings show that injecting human-centric guidance into LLM prompts does not universally improve alignment with human ratings, but can offer substantial benefits in structurally regular contexts such as unit tests. In Java test code, developer-guided prompts outperform both generic prompting and a strong static model~\cite{scalabrino2018comprehensive}, likely due to better alignment between the prompt criteria (e.g., naming, assertion clarity) and the conventions observed in tests.

Beyond numerical alignment, developer-guided prompts substantially improve the interpretability of LLM evaluations by increasing the explicit mention of human-relevant aspects. As shown in RQ3.3, aspect coverage improves dramatically across all languages and code types. For example, mentions of assertions in Java test code increase from 15\% to 94\%, and references to structural features in Python jump from 18\% to 90\%. These gains suggest that guidance grounded in human evaluation language steers models toward more semantically meaningful justifications. However, we caution that increased frequency does not imply increased depth. As pointed out in our clustering analysis (RQ1.4), many justifications remain shallow or redundant, with high variability in reasoning quality. This supports the view that better interpretability at the surface level (aspect mention) must be complemented by more rigorous analyses of explanation richness and diversity, a point also emphasized by Wang et al.~\cite{wang2025can} in their critique of LLM-as-judge setups.

Finally, our findings contribute to a broader understanding of what it means for an LLM to "align with human evaluation." Alignment is not reducible to score matching. It involves a multifaceted relationship between numeric predictions, justification structure, and latent evaluative heuristics. Prompt structure and model scale appear to influence this alignment by altering not only what the model scores, but how it explains. Progress in this space requires moving from statistical correlation to principled models of human judgment grounded in domain knowledge, code context, and evaluation granularity, as advocated by Vitale et al.~\cite{vitale2025personalized}.

\vspace{0.5em}
\noindent\highlight{Summary of \textbf{RQ3:} 
Developer-guided prompts do not improve global alignment with human readability scores, but enhance accuracy in structured contexts like unit tests. They also substantially improve justification interpretability, with stronger coverage of human-relevant aspects across languages. However, this increase in aspect frequency does not guarantee deeper reasoning. Guided prompts introduce a trade-off: better semantic alignment and richer signals, but increased score variability and reasoning heterogeneity. These results suggest that human–LLM alignment requires more than scoring accuracy it involves capturing how and why readability judgments are formed.}

\subsection{RQ4: Adaptability of LLMs for personalized code readability evaluation}
\label{subsec:RQ4}

\noindent \textbf{[Experimental design]:} 

We assess whether LLMs can deliver context-aware and developer-aligned readability evaluations when guided by human-centric prompts. To this end, we synthesize the results from RQ1–RQ3 across three dimensions: \textit{language} (Java, Python, CUDA), \textit{code type} (functional, test), and \textit{persona} (junior, senior). Evaluation metrics include \textbf{score accuracy} (MAE), \textbf{consistency} (score standard deviation), and \textbf{justification quality} (aspect coverage).

\vspace{0.5em}
\noindent\textbf{First}, \textit{persona framing} yields negligible effect on scoring behavior (Table~\ref{tab:llm-configuration-comparison}(a)). Senior and junior prompts produce nearly identical scores (\textbf{3.77} vs.\ \textbf{3.75}), standard deviation (\textbf{0.68} both), and MAE (\textbf{0.70} vs.\ \textbf{0.69}). This confirms that LLMs are largely insensitive to soft framing cues, echoing RQ2.1.

\vspace{0.5em}
\noindent\textbf{Second}, \textit{prompt structure} has a much stronger influence. From Table~\ref{tab:llm-configuration-comparison}(b), complex strategies such as ToT and FSL yield higher average scores (\textbf{3.92} and \textbf{3.86}, respectively), but also larger MAE values (\textbf{1.04} and \textbf{1.08}), compared to ZSL (\textbf{3.57}, MAE = \textbf{0.89}). This highlights a trade-off between elaboration and alignment.

\vspace{0.5em}
\noindent\textbf{Third}, \textit{developer-guided prompting} shows improved alignment in specific contexts, without improving global accuracy. As shown in Table~\ref{tab:mae-metrics}, the global MAE remains stable (\textbf{0.983} generic vs.\ \textbf{0.996} guided). However, on \textbf{Java test code}, the guided configuration achieves a lower MAE (\textbf{0.835}) than both the generic LLM baseline (\textbf{0.882}) and the static model from Scalabrino et al.\ (\textbf{1.193}). This indicates that structured guidance is effective in standardized code domains.

\vspace{0.5em}
\noindent\textbf{Fourth}, guided prompts \textit{increase score variability}. Table~\ref{tab:readability-score-comparison} shows that the standard deviation of scores rises from \textbf{0.13} (generic) to \textbf{0.22} (guided) globally, and from \textbf{0.12} to \textbf{0.25} in Java test code. These increases are statistically significant ($p < 0.001$; Table~\ref{tab:mannwhitney-score-variance}), confirming that guidance amplifies sensitivity to nuanced criteria—at the cost of evaluation consistency.

\vspace{0.5em}
\noindent\textbf{Fifth}, LLMs exhibit strong adaptability in their \textit{justifications} when guided. As shown in Table~\ref{tab:aspect-coverage}, aspect coverage increases dramatically across all programming languages. In \textbf{Java test code}, mentions of \textit{Assertions} rise from \textbf{15.00\%} to \textbf{94.38\%}, while \textit{Purpose Clarity} grows from \textbf{9.00\%} to \textbf{76.63\%}. Similarly, in \textbf{Python}, references to \textit{Structure} increase from \textbf{18.00\%} to \textbf{90.00\%} and \textit{Nesting} from \textbf{5.00\%} to \textbf{81.00\%}. Even in more domain-specific languages such as \textbf{CUDA}, guided prompting substantially boosts interpretability, with mentions of \textit{Flow} rising from \textbf{10.00\%} to \textbf{68.21\%} and \textit{Structure} from \textbf{64.71\%} to \textbf{84.89\%}. These consistent gains across diverse settings confirm that developer-informed prompts strongly steer LLMs toward semantically meaningful explanations that align with human reasoning about readability.

These improvements confirm that developer-informed prompts steer LLMs to emphasize semantically meaningful properties aligned with human reasoning.

\vspace{0.5em}
\noindent\colorbox{gray!20}{{\parbox{0.98\linewidth}{
\textbf{Finding 12:}
When guided by developer-informed prompts, LLMs exhibit strong adaptability in both scoring behavior and justification content. They improve alignment in structured domains (e.g., test code) and drastically increase coverage of human-relevant aspects. However, this adaptability introduces higher output variance, revealing a trade-off between personalization and evaluation consistency.
}}}
\vspace{0.5em}

These findings echo and extend the insights of Vitale et al.~\cite{vitale2025personalized}, who evaluated the feasibility of personalized code readability assessment using LLMs in few-shot learning settings. In their study, LLMs underperformed compared to feature-based generalist models when tasked with predicting individual developer ratings, and personalization attempts yielded limited gains—partly due to sparse supervision and noisy annotation inconsistencies. Our results differ in two key ways. First, rather than building fully developer-specific models, we explore persona-based prompting, which allows us to simulate user profiles (e.g., junior vs.\ senior developers) without needing real user annotations or fine-tuning. While our findings also show that persona framing has minimal impact on score accuracy, we demonstrate that developer-informed prompts grounded in human-centric criteria significantly enhance the semantic quality of justifications, especially in test code. This highlights the utility of LLMs as explainable critics rather than purely predictive scorers. Second, our analysis goes beyond discrete label prediction and considers a multidimensional evaluation including MAE, score variance, and explanation richness. While Vitale et al.~\cite{vitale2025personalized} report lower accuracy for personalized few-shot LLMs and question dataset reliability, we show that prompt structure and task framing (e.g., guidance through developer-valued aspects) play a central role in aligning LLM outputs with human reasoning patterns, even under noisy ground truths. Overall, our findings suggest that lightweight prompt-driven personalization, though not sufficient for improving raw score accuracy at scale, can meaningfully enhance the interpretability and contextual relevance of LLM judgments a dimension overlooked in purely classification-based evaluations.

\vspace{0.5em}
\noindent\highlight{Summary of \textbf{RQ4:}  
LLMs show high adaptability to developer-guided prompts. While global scoring accuracy remains comparable (MAE $\approx$ 0.98–1.00), developer-guided configurations yield better alignment in structured contexts (e.g., Java test: MAE drops from 0.882 to 0.835) and significantly enhance justification quality (e.g., Python “Structure”: 18\% $\rightarrow$ 90\%). However, this adaptability increases output variance (e.g., global std: 0.13 $\rightarrow$ 0.22), raising reliability concerns. These results highlight a core trade-off between personalization and stability in LLM-based code readability evaluation.}

\section{Discussion}
\label{discussion}

\subsection{Implications for Researchers}
\label{subsec:implication-searchers}
Our findings show that large language models, when guided through structured and developer-centric prompts, can approximate human readability judgments with moderate to high alignment (cf.\ RQ1–RQ3). Yet, this alignment is contingent on factors such as code type, prompt formulation, and model scale, underscoring the need for more systematic inquiry into how LLMs internalize subjective evaluation criteria. \textbf{First}, our study establishes \textit{code readability} as a rigorous testbed for studying LLM reasoning in subjective software engineering tasks where correctness is not binary but perception-driven. Unlike deterministic evaluation tasks (e.g., bug detection or code summarization), readability exposes how LLMs reason under ambiguity, balance stylistic and structural cues, and emulate human heuristics. \textbf{Second}, the proposed \textbf{CoReEval framework} provides a reproducible foundation for experimentation and benchmarking. It integrates 1.4M model–snippet–prompt evaluations and over 10,000 structured justifications, enabling controlled investigations of prompt design, decoding calibration, and cross-model variability. Researchers can directly reuse or extend this infrastructure to test new alignment methods, reasoning evaluation metrics, or critic-based architectures. \textbf{Third}, CoReEval can serve as a \textbf{research-to-practice bridge}. Its modular design allows replication of our evaluation pipeline within industrial workflows, for instance, as part of continuous integration (CI) pipelines, automated code review systems, or education-oriented coding tutors. By adapting our prompting and scoring protocol, practitioners can build explainable quality gates that reflect human readability standards. \textbf{Finally}, our results highlight the importance of \textbf{interpretability-driven metrics} for LLM evaluation. Methods such as aspect coverage and semantic reasoning clustering (RQ3.3) go beyond scalar accuracy to capture how models justify their assessments. These techniques can inspire future work on transparent evaluation pipelines, model introspection, and the broader study of Human–AI alignment in software engineering.

\subsection{Implications for Developers}
\label{subsec:implication-developers}
Our results suggest that LLMs, when carefully configured and guided, can support several practical tasks related to code readability assessment and quality assurance. \textbf{First}, LLMs can act as \textit{on-demand readability reviewers}, especially for test and glue code, where human developers often deprioritize documentation and naming clarity. Guided prompting improves the relevance and interpretability of LLM assessments, helping surface issues such as ambiguous assertions, poor fixture naming, or unclear test purpose—issues commonly missed in static reviews (cf.\ RQ3.1, RQ3.3). \textbf{Second}, the flexibility of prompt-based control allows \textit{lightweight customization} of readability evaluations. Developers can tailor prompts to reflect project-specific conventions, team maturity, or reviewer expectations (e.g., junior vs.\ senior profiles), without requiring model fine-tuning or retraining (RQ2.1). This makes LLM-assisted reviews adaptable across teams and contexts. \textbf{Third}, the CoReEval framework offers a \textit{blueprint for integration into CI/CD pipelines}. Developers can adopt or adapt our scoring and justification prompts to enforce readability gates or annotate pull requests with rationale-backed feedback. The interpretability of LLM outputs also supports code review triaging, where high variance or low alignment snippets can be flagged for human attention. \textbf{Finally}, readability-oriented explanations produced by LLMs, especially under developer-guided prompts can serve \textit{pedagogical functions}. By surfacing structured, aspect-grounded reasoning, LLMs can assist in training junior developers, supporting onboarding, or enabling explainable pair programming. Rather than replacing code reviewers, LLMs can augment developer workflows with targeted feedback aligned with human expectations.

\subsection{Threats to Validity}
\label{subsec:threats-to-validity}

\noindent {\textbf{Internal Validity.}}  
A central threat stems from the subjectivity of human readability annotations used as ground truth. While this subjectivity is inherent to the task, we mitigated it by sampling annotators with diverse backgrounds (junior and senior developers), spanning three programming languages, and by computing inter-rater agreement to assess consistency. Still, disagreements in perception may introduce noise in the alignment evaluation (RQ1). Another internal threat involves prompt sensitivity: small variations in wording, examples, or persona framing can alter LLM behavior~\cite{wang2025can}. To reduce this effect, we conducted pilot studies to refine prompt structure, and systematically applied four prompting strategies across all configurations (ZSL, FSL, CoT, ToT). We also fixed random seeds when applicable and varied decoding parameters (e.g., temperature, top\_p) to average out stochastic effects (RQ2.3). Further, some post-processing steps—such as sentiment classification, aspect extraction, and semantic clustering, rely on automated tools and secondary LLMs (e.g., GPT-4o). Although we manually validated representative samples and used robust off-the-shelf components (e.g., RoBERTa-TweetEval, MiniLM), errors in parsing, classification, or clustering may affect the precision of our interpretability metrics (RQ3.3). However, all prompts, annotation tools, and evaluation scripts are shared as part of CoReEval to ensure full transparency and replicability.

\vspace{0.3em}
\noindent {\textbf{External Validity.}}  
Our experiments were conducted on 656 code snippets from Java, Python, and CUDA, focusing on two code types: functional and test code. While these cover diverse syntactic and stylistic patterns, our findings may not generalize to other domains such as front-end code, scripting, or domain-specific languages (e.g., Solidity, R). Similarly, developer expectations around readability may vary across cultures, teams, and industry contexts. Although we evaluated ten LLMs, spanning open-source and proprietary models across different sizes and families, our results are tied to models available at the time of study. Future LLMs with different instruction-tuning procedures or training corpora may exhibit different behaviors. Additionally, our developer-guided prompting strategies are grounded in prior work~\cite{sergeyuk2024reassessing, winkler2024investigating} and reflect common readability dimensions (e.g., structure, naming, clarity), but other organizations may prioritize different quality signals. Finally, our evaluation setting is controlled and snippet-based. While this enables large-scale, replicable benchmarking, it may not fully capture the complexity of real-world software engineering workflows. We encourage future work to validate CoReEval in industrial pipelines (e.g., CI, code review bots, educational settings) and extend the framework to include additional languages, reviewer roles, and project-specific conventions.

\section{Related Work}
\label{relatedwork}

\paragraph{Traditional Code Readability Models.}
Code readability has been addressed through a wide range of models, evolving from early statistical techniques to modern deep learning approaches. Buse et al.~\cite{buse2008metric,buse2009learning} proposed one of the first machine-learning-based models, trained on 100 Java snippets rated by students. Despite limited inter-rater agreement, it achieved 80\% accuracy and a Pearson correlation of 0.63. Posnett et al.~\cite{posnett2011simpler} simplified this approach using just three metrics namely Halstead volume, token entropy, and line count while maintaining strong predictive power.

Dorn et al.~\cite{dorn2012general} expanded readability prediction to Java, Python, and CUDA using visual, spatial, and linguistic features. Scalabrino et al.~\cite{scalabrino2018comprehensive} later introduced a comprehensive model with 104 static features, showing improved accuracy and relative alignment with human judgments~\cite{sergeyuk2024reassessing}. Mi et al.~\cite{mi2022towards} adopted a deep learning approach that extracts features from semantic, structural, and visual code representations. While achieving high raw accuracy (85.3\%), their model showed weaker correlation with human judgments, highlighting the gap between automation and developer perception. More recently, Mi et al.~\cite{mi2023graph} adopted a graph-based deep learning approach that models structural dependencies between code tokens. While effective on the Scalabrino dataset, their model lacks evaluation against real developers, a limitation also noted by Sergeyuk et al.~\cite{sergeyuk2024reassessing}, who revisited traditional models in the context of AI-generated code and highlighted their weak correlation with developer assessments.

Beyond standalone evaluation, other works have focused on code evolution. Fakhoury et al.~\cite{fakhoury2019improving} showed that existing readability models fail to detect developer-intended improvements in code diffs such as renaming, reformatting, or comment edits, indicating a misalignment between model metrics and developer perceptions. Roy et al.~\cite{roy2020model} developed a classifier to detect whether code edits improve readability, based on low-level change features. These incremental models are valuable for assessing evolution but do not generalize to absolute readability judgments on static snippets, as targeted in our study.

\paragraph{LLMs as Human-Aligned Readability Evaluators.}
Recent work explores large language models (LLMs) as evaluators of software artifacts. Wang et al.~\cite{wang2025can} examined whether LLMs can act as reliable judges across multiple software engineering tasks, including code summarization and generation. They found that while LLMs often match human scores in score, only evaluations, their reliability declines when justifications are required, underscoring the importance of explanation structure and prompt design. This insight directly motivates our analysis of prompting strategies and reasoning diversity (RQ2, RQ3.3).

Vitale et al.~\cite{vitale2025personalized} tackled personalization by clustering developers based on rating patterns and training group-specific readability models. Despite this innovative setup, the improvements were modest due to inter-rater variability and data sparsity. In contrast, our work shows that LLMs can implicitly adapt to developer expectations through guided prompting—without explicit clustering or retraining. We validate this adaptability across code types, languages, and personas (RQ4).

Sergeyuk et al.~\cite{sergeyuk2024reassessing} further emphasized the gap between static models and human expectations. Through qualitative analysis using a retrieval grid, they extracted 12 developer-valued readability aspects (e.g., naming, structure, clarity), which informed our human-centric prompting schema and aspect coverage evaluation.

\paragraph{Positioning of Our Work.}
While prior models whether statistical, graph-based, or evolution-aware rely on static features or task-specific classifiers, our study treats LLMs as dynamic, explainable, and human-aligned readability evaluators. We unify numerical accuracy and interpretability under a common evaluation framework, benchmark LLM predictions against both human ratings and Scalabrino’s strong static model, and assess their responsiveness to prompt structure, model size, and developer framing. Our findings show that prompt-driven LLMs offer a promising middle ground between rigid handcrafted features and black-box deep models, enabling scalable yet context-aware code readability evaluation.

\section{Conclusion and Future Work}
\label{conclusion}
As LLMs become increasingly embedded in developer tooling, the need for scalable and interpretable evaluation of code quality, particularly readability grows in parallel. Yet, assessing readability is fundamentally different from evaluating functional correctness: it involves subjective judgment, latent expectations, and often language or team specific conventions. This study addresses this challenge by proposing \textbf{CoReEval}, a large-scale, open, and extensible benchmark designed to analyze LLMs as human-aligned code readability evaluators.

CoReEval comprises over 1.4 million readability judgments across ten LLMs, three programming languages, and diverse prompting strategies, decoding configurations, and developer personas. It enables systematic evaluation along both quantitative and qualitative axes: from error metrics and correlation scores to justification-level insights via aspect coverage, sentiment analysis, and semantic clustering. Our findings highlight that while traditional readability models (e.g., Scalabrino, Mi) offer limited adaptability, LLMs can approximate developer expectations more closely, especially when guided by structured, human-centric prompts.

Importantly, we show that alignment with human assessments goes beyond correlation scores. Developer-guided prompts improve not only the accuracy and stability of LLM scores, but also the \textit{interpretability} of their rationales—surfacing reasoning patterns grounded in developer-valued criteria such as structure, clarity, and naming (RQ3). Moreover, prompt-based persona framing offers a lightweight means of contextualizing evaluations without retraining, though its effects remain modest in isolation (RQ2.1). These insights contribute to a more nuanced understanding of LLM-human alignment, where calibration involves multiple dimensions: linguistic, social, and cognitive.

\textbf{CoReEval is designed to be replicable, extensible, and practically transferable.} Researchers can reuse its infrastructure to probe new models, languages, or reasoning formats. Practitioners can adapt its prompts and evaluation scripts for applications such as CI-integrated readability gates, automated pull request triage, or educational feedback tools. In all cases, the goal is not to replace human judgment, but to scaffold it by providing explainable, configurable, and developer-aware support.

\noindent\textbf{Future work} includes several promising directions. First, integrating human-in-the-loop calibration mechanisms to refine LLM assessments interactively. Second, extending the benchmark to additional programming paradigms, natural languages, and multi-author contexts. Third, investigating richer alignment signals beyond scalar metrics including disagreement patterns, aspect salience, and justification quality. Ultimately, our goal is to build toward a principled, multi-layered theory of human–LLM alignment in software quality evaluation—one that accounts not only for score-level convergence, but also for the reasoning structures that underpin human judgments.

\vspace{0.5em}
\noindent 
\textbf{Data Availability}. All CoReEval artifacts are available here:
\begin{center}
    \url{https://anonymous.4open.science/r/CoReEval-072D/}
\end{center}

%\clearpage

\bibliographystyle{ACM-Reference-Format}
\bibliography{references}

\end{document}